\newif\iffull
\newif\ifsoda
\newif\ifTALG
\newif\ifARXIV
\newif\ifrunningonARXIVserver
\newif\ifjustsyntaxchecking
\fullfalse
\fulltrue
\TALGtrue
\ARXIVtrue
\runningonARXIVservertrue 

\ifARXIV\TALGtrue\fi

\ifsoda
\documentclass[twoside,leqno,twocolumn]{article}
\usepackage{ltexpprt}
\else
\ifTALG
\ifjustsyntaxchecking
\documentclass[acmsmall,nonacm]{acmart}
\else
\ifARXIV
\documentclass[acmsmall,
nonacm, 
screen
]{acmartrote}
\acmJournal{TALG}
\else
\documentclass[acmsmall,
screen,timestamp]{acmart}
\fi
\acmJournal{TALG}
\fi
\else
\documentclass[a4paper]{article}
\usepackage[top=2cm,bottom=2.5cm]{geometry}
\fi
\fi

\ifARXIV\else
\usepackage{linenum}\linenumberseparation=25pt

\fi

\ifjustsyntaxchecking
\input prepareForSyntaxChecking
\fi

\usepackage{amsmath}

\ifsoda
\makeatletter
\let\@old@endproof=\@endproof
\def\@endproof{\qed\@old@endproof}
\makeatother

\newcommand{\openbox}{\leavevmode
  \hbox to.77778em{%
  \hfil\vrule
  \vbox to.675em{\hrule width.6em\vfil\hrule}%
  \vrule\hfil}}
\providecommand{\qedsymbol}{\openbox}
\DeclareRobustCommand{\qed}{%
  \ifmmode \mathqed
  \else
    \leavevmode\unskip\penalty9999 \hbox{}\nobreak\hfill
    \quad\hbox{\qedsymbol}%
  \fi
}
\fi

\usepackage{amsfonts}
\usepackage{graphicx}
\usepackage{url}
\usepackage{hyperref}
\usepackage{tikz}
\usepackage{gnuplot-lua-tikz}

\usepackage{enumitem}
\newtheorem{observation}{Observation}

\newcommand\reals{\mathbb{R}}
\newcommand\mds{minimal dominating set}
\newcommand\MDS{Minimal Dominating Set}
\newcommand\wstree{well-structured enumeration tree}
\newcommand\cwstree{\wstree}
\newcommand\first{first}
\newcommand\mes{\textit{message}}
\newcommand\sta{\textit{state}}
\newcommand\child{\textit{child}}
\newcommand\from{\leftarrow}

\ifsoda\else
\ifjustsyntaxchecking
\else
\fi
\fi

\newcommand{\commandb}[1]{%
\valign{##\cr
\vskip 3pt
\hbox{\begin{tabular}{rl}
\omit\hskip 2,6cm\null&\omit\hskip 4,cm\\
#1
\end{tabular}}%
\vskip 1pt
\cr
\noalign{\kern -0,5mm\hskip - 4,cm}
\cleaders\vbox to 2pt{\vss\hbox{\small.}\vss}\vfill\cr
}\hskip 4,cm
}

\newcommand{\commandbox}[2]{%
\vbox{\hbox{#1}
\vskip 3pt
\hbox{\fbox{%
\commandb{#2}}}}}

\DeclareMathOperator\conv{conv}
\DeclareMathOperator\hull{hull}
\def\hullp{\mathop{\hull^+}}
\begin{document}

\ifsoda
\else
\ifTALG
\title
[\MDS s in a Tree: Counting, 
  Enumeration, and Extremal Results]
{\MDS s in a Tree:\\ Counting, 
  Enumeration, and Extremal Results
}
\else
\title{\MDS s in a Tree:\\ Counting, 
  Enumeration, and Extremal Results}
\fi
\fi
\ifTALG
\author{G\"unter Rote}
\affiliation
{%
  \institution{Freie Universit\"at Berlin}
\department{Institut f\"ur Informatik}
\streetaddress{Taku\-stra\ss e~9}
\postcode{14195}
\city{Berlin}
\country{Germany}}
\email{rote@inf.fu-berlin.de}
\orcid{0000-0002-0351-5945}

\acmPrice{250.00}
\acmDOI{}
\citestyle{acmauthoryear}

\ifARXIV\else
\ccsdesc[500]{Mathematics of computing~Enumeration}
\ccsdesc[300]{Mathematics of computing~Graph enumeration}
\fi

\else
\author{ G\"unter Rote\thanks{Institut f\"ur Informatik, Freie
    Universit\"at Berlin, Taku\-stra\ss e~9, 14195 Berlin, Germany,
    \href{mailto:rote@inf.fu-berlin.de}{\texttt{rote@inf.fu-berlin.de}}}}
\fi

\begin{abstract}
  A tree with $n$ vertices has at most $95^{n/13}$ minimal dominating
  sets.
The growth constant $\lambda = \sqrt[13]{95} \approx 1.4194908$
is best possible.
It is obtained in a semi-automatic way as a kind of ``dominant eigenvalue'' of a bilinear
operation on sixtuples that is derived from the dynamic-programming
recursion for computing the number of \mds s of a tree.
We also derive an output-sensitive algorithm for listing all \mds s with
linear set-up time and linear delay between successive solutions.
\end{abstract}

\maketitle

\newif\ifINTOTOC
\makeatletter 
\def\l@section#1#2{\relax \ifnum 1>\c@tocdepth \else
  \global\INTOTOCtrue
  {\def\hyper@linkstart ##1##2##3\hyper@linkend{%
      \def\habat{##3}
      \def\habab{Acknowledgments}\ifx\habab\habat\global\INTOTOCfalse\fi
      \def\habab{Abstract}\ifx\habab\habat\global\INTOTOCfalse\fi
      \def\habab{Contents}\ifx\habab\habat\global\INTOTOCfalse\fi
    }
    \def\habat{#1}
    \def\habab{{Acknowledgments}}\ifx\habab\habat\global\INTOTOCfalse\fi
    \def\habab{{Abstract}}\ifx\habab\habat\global\INTOTOCfalse\fi
    \def\habab{{Contents}}\ifx\habab\habat\global\INTOTOCfalse\fi
    \ifrunningonARXIVserver\else 
    #1\fi}
  \ifINTOTOC  
  \par \addpenalty
  \@secpenalty \addvspace {0.25em}%
  \begingroup \hyphenpenalty \@M \@ifempty
 {2pc}{\@tempdima \csname r@tocindent\number 1\endcsname \relax
 }{\@tempdima 2pc\relax }\parindent \z@ \leftskip 1pc\relax \advance
 \leftskip \@tempdima \relax \rightskip \@pnumwidth
 plus4em \parfillskip -\@pnumwidth \leavevmode \hskip -\@tempdima
 #1\nobreak \relax \hfil \hbox to\@pnumwidth {\@tocpagenum {#2}}\par
 \nobreak \endgroup \fi\fi }
\makeatother

\ifsoda\else
\tableofcontents

\fi

\pagestyle{standardpagestyle}

\section{Introduction}


\paragraph{Problem Statement.}
 A vertex $a$ in an undirected graph $G=(V,E)$
\emph{dominates} a vertex $b$ if $b=a$ or $b$ is adjacent to $a$.
A \emph{dominating set} in a graph $G=(V,E)$ is a subset $D\subseteq
V$ such that every vertex is dominated by some element of $D$.
In other words, every vertex $a\in V-D$ must have a neighbor in
$D$.
$D$ is
a \emph{minimal dominating set} if no proper subset of $D$ is
a dominating set.
A more concrete characterization of \mds s is a follows.
A dominating set $D$ is a \mds\ iff every vertex $a\in D$ has a
\emph{private neighbor}: a vertex $b$ that
dominated by $a$ but by no other vertex in $D$. (The private ``neighbor''
can be the vertex $a$ itself.)

\paragraph{Results.}
Let $M_n$ denote the maximum number 
 of \mds s that a tree with $n$
vertices can have.
 We provide the correct and tight value of the
growth constant $\lambda$ of $M_n$.


\begin{theorem}\label{sizes}
Let $\lambda = \sqrt[13]{95} \approx 1.4194908$.

\begin{enumerate}
\item 
A tree with $n$ vertices has at most 
$2 \lambda^{n-2}< 0.992579 \cdot \lambda^n$ \mds s.
\item 
  For every $n$, there is a tree with
at least $0.649748 \cdot \lambda^n$ \mds s.
\item 
  For every $n$ of the form $n=13k+1$, there is a tree with
at least $95^k > 0.704477 \cdot \lambda^n$ \mds s.
\end{enumerate}
\end{theorem}

On the algorithmic side, we derive an output-sensitive algorithm for
enumerating all solutions:

\begin{theorem}
\label{theorem-enumeration}
The \mds s of
  a tree with $n$ vertices can be enumerated with $O(n)$ setup time
and with $O(n)$ delay between successive solutions.
\end{theorem}

\paragraph{Previous Results.}

Marcin Krzywkowski
\shortcite{k-tmmds-13}
gave
an algorithm for
listing all minimal dominating sets of a tree of order $n$ in
time $O (1.4656^n)$, thus proving that every tree has at most
$1.4656^n$ minimal dominating sets.
 Golovach, Heggernes, Kant\'e, Kratsch and
 Villanger~\shortcite{ghkkv-mdsig-17}
recently improved this upper bound to $3^{n/3}\approx 1.4422^n$.

\begin{figure}[bth]
  \centering
  \includegraphics{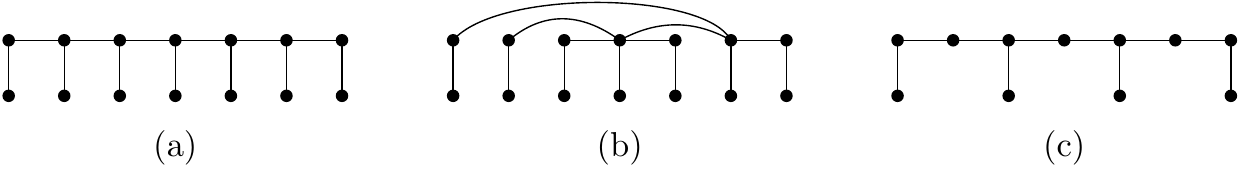}
  \Description{The comb is a path with a lead added to each
    vertex. The generalized comb is described in the text. The
    extended comb in part (c) is described in Section~\ref{sec:optimize}.}
  \caption{(a) the comb graph with 7 teeth, (b) a generalized comb, (c) an extended comb.}
  \label{fig:comb}
\end{figure}

Small examples indicate that the class of \emph{comb graphs} of
Figure~\ref{fig:comb}a
with an even number $n$ of vertices and $n/2$ teeth might have the
largest number of \mds s. They have $2^{n/2}\approx 1.4142^n$ \mds s,
because one can independently choose a vertex out of every tooth
(see Observation~\ref{obse}(\ref{leaf-condition}) below).
The class of graphs with so many \mds s is in fact very large:
One can take \emph{any} tree on $n/2$ vertices and append a leaf to each
vertex, as in
Figure~\ref{fig:comb}b.
The trees with odd $n$ seem to have much fewer than
$1.4142^n$ \mds s.
It turns out that
these observations are indeed true for $n\le 18$, but they fail for
larger $n$, see Figure~\ref{fig:tab} and Table~\ref{result}
in Section~\ref{upper1}.

The best lower bound on the growth constant
$\lambda$ that has been known so far
is $\sqrt[27]{12161}\approx 1.416756$, due to
Krzywkowski~\shortcite{k-tmmds-13}.
Krzywkowski constructed a tree with 27 vertices and
12161 \mds s. Since the sequence $M_n$ is supermultiplicative
(Observation~\ref{obse}(\ref{super-function}) below), this establishes
$\sqrt[27]{12161}$ as a
lower bound on $\lambda$.

It occurs frequently in combinatorics that a lower bound is
established through a particular example, from which the asymptotic growth is
 derived with the help of supermultiplicativity.  However, in
our case, this method is bound to fail in finding the true lower
bound: By Part~1 of Theorem~\ref{sizes}, a tree with $n$
vertices that would have $\lambda^n$ \mds s does not exist.
By contrast, our lower bound $\lim \sqrt[n]{M_n} \ge \lambda$
will be established by
an infinite family
of trees (Section~\ref{lower}).

The question can of course be asked for other graph classes than trees, and there is an
extensive literature, see 
\cite{COUTURIER201382} for an overview.
On general graphs, the best upper bound is $1.7159^n$, and no graph
with
$n$ vertices and
more than $1.5705^n$ minimal dominating sets is known.

\paragraph{Techniques.}
While we settle the question of the growth constant for trees,
we believe that the techniques that lead to the result are more
interesting than the result itself.

We start with a standard dynamic-programming algorithm for counting
the number of \mds s of a \emph{particular} tree
(Section~\ref{counting}).
The algorithm operates on sixtuples of numbers, because there
happen to be
six classes of partial solutions that must be distinguished.
We then abstract the calculation from a particular tree, and
deduce an algorithm for finding all sixtuples that can arise for
a particular number $n$ of vertices. From this, it is easy to calculate~$M_n$.

Finally, we will try to enclose the set of sixtuples in a
six-dimensional geometric body. If we succeed to find an appropriate
shape with certain properties, which depend on some
putative value of $\lambda$, we have established $\lambda$ as an upper
bound of the growth constant
(Proposition~\ref{characterize} in Section~\ref{upper2}).
This suggests a
semi-automatic computer-assisted method for searching for the correct
 growth constant 
 (Section~\ref{automatic}).

As a side result, our dynamic-programming setup can be adapted to an
efficient
enumeration algorithm for listing all \mds s of a tree
(Theorem~\ref{theorem-enumeration}) with linear delay,
see Section~\ref{listing}. Previous algorithms
\cite{k-tmmds-13,ghkkv-mdsig-17}
were not even output-sensitive in the sense of being polynomial in the
combined size of the input and output.

Our results were presented in preliminary form at the
ACM--SIAM Symposium on Discrete Algorithms (SODA19)
in San Diego in
January 
2019~\cite{r-mnmds-19}.

\section{Preliminaries}
\label{Preliminaries}
\def\definitionlegal
{Let $D\subseteq V$ be a set of vertices in a graph $G$.
It is useful to rephrase the conditions for \mds s:
  We call a
vertex $a\in V$ \emph{legal} if
\begin{enumerate}
\item [(a)] $a\in D$ and $a$ has a private neighbor, or
\item [(b)] $a\notin D$ and $a$ is dominated, i.e., it has some
  neighbor in $D$.
\end{enumerate}
Thus, $D$ is a \mds\ iff all vertices of the graph are legal.
}

\iffull
\definitionlegal

We will now
establish the well-known fact that the numbers $M_n$ are
supermultiplicative.
\else
We will make a simple observation and mention the well-known fact that the numbers $M_n$ are
supermultiplicative.
\fi

\begin{observation}\label{obse}
  \begin{enumerate}
  \item  \label{leaf-condition}
  If $a$ is leaf and $b$ its neighbor, then every \mds\ $D$ contains
  exactly one of $a$ and $b$.
Moreover, $a$ can always be chosen as the private neighbor of this
vertex.
\iffull
\item \label{twins}
  If $a_1,\ldots,a_k$ are leaves with a common neighbor $b$, then
either all vertices $a_1,\ldots,a_k$ belong to $D$ or none of them
 belongs to $D$.
(We will call two leaves that have a common neighbor \emph{twins}.)
\item \label{super-trees}
If $T_1$ and $T_2$ are two trees with
 $M(T_1)$
and
 $M(T_2)$
\mds s, there is a way to insert an edge between 
 $T_1$ and $T_2$ such that the resulting tree has
exactly
 $M(T_1)M(T_2)$ \mds s, except when
 $T_1$ and $T_2$ are two singleton trees.
\fi
\item \label{super-function}
The function $M_n$ is supermultiplicative:
$$M_{i+j}\ge M_{i}M_j
$$
for $i,j\ge 1$.
  \end{enumerate}
\end{observation}
\def\proofofobservation
{
\begin{proof}
  Statement 1 is easy to see, and Statement 2 follows directly from it.

  For the third claim, consider first the case that both $T_1$ and
  $T_2$ have at least 2 vertices. Let $a_i$ be a leaf in $T_i$ and $b_i$ be its
  neighbor.  Then we connect the trees by the edge
  $b_1b_2$. We argue that
the presence of this edge 
makes no difference for the \mds s in the union of the two trees.
An edge $b_1b_2$ could in principle affect the legality of
$b_1$ or $b_2$ or a neighbor of $b_1$ or $b_2$.
However, (i) $b_1$ is always dominated either by $a_1$ or by $b_1$, no
matter whether the edge $b_1b_2$ is present.
(ii) Independently of whether we choose $a_1$ or $b_1$ as an element
of $D$ or not, we can
always choose $a_1$ as a private neighbor for it;
the edge $b_1b_2$ is not required to find a private neighbor.
(iii) $b_1$ can never be
used as a
private neighbor of another vertex than $a_1$ or $b_1$ because it
 is already
  dominated by $a$ or $b$.
Thus the presence or removal of $b_1b_2$ will neither help nor prevent any vertex to
find a private neighbor.



When one of the trees, say $T_1$, is a singleton tree, we connect it to
a neighbor $b_2$ of a leaf $a_2$ in $T_2$. In the resulting tree,
$a_2$ has a new twin, and thus $M(T_2)$ is unchanged. In view of
$M(T_1)=1$, this is what we need.


Supermultiplicativity in the fourth claim follows from Statement~3.
 The exceptional case $i=j=1$, when $T_1$ and $T_2$ are two singleton trees,
 can be checked directly.
\end{proof}
}
\ifsoda
\else 
\proofofobservation
\fi

\begin{figure}[htb]
  \centering
  \ifsoda
\else
  \includegraphics[scale=1]{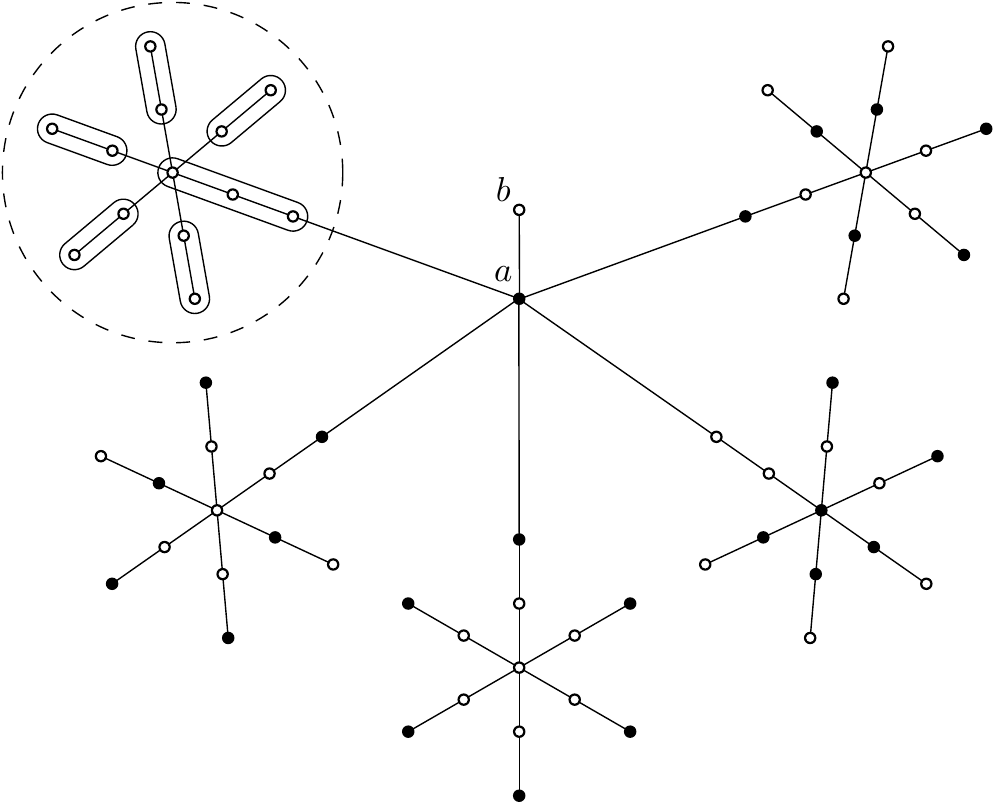}
  \fi
  \Description{Illustration of a \mds $D$ in a star of snowflakes, as
    described in the text}
  \caption{A star of 5 snowflakes. The vertices of $D$ are black.}
  \label{snowflake}
\end{figure}
\section{Lower Bound Example: the Star of Snowflakes}
\label{lower}

The lower bound on the constant $\lambda$ is proved by
the \emph{star of snowflakes} (Figure~\ref{snowflake}),
 a family of examples with $13k+2$ vertices and at least
$95^k$ \mds s, for $k\ge1$.
Through the analysis of this example, we hope that the reader may get
familiar with \mds s.
A single snowflake has 13 vertices and consists of 6 paths of two edges each,
attached to a central vertex.
We take the union of $k$ snowflakes and 
a separate \emph{root vertex} $a$, and we connect $a$ to a leaf of
each snowflake.
In addition, $a$ gets another leaf $b$ as a neighbor,
for a total of $13k+2$ vertices.
Let us count the \mds s containing $a$.
We will first check that 95 possibilities can be independently chosen in each snowflake:
We  
partition each snowflake into five groups of size 2 and one group of
size 3, as shown
\iftrue
in the snowflake
\fi
at the top left of
Figure~\ref{snowflake}. It is now straightforward to check that a \mds\
 must contain exactly one vertex from each group.
(For the five groups of size 2, this follows directly from Observation~%
\ref{obse}(\ref{leaf-condition}).)
Out of these $3\cdot2^5=96$ possibilities, one possibility is forbidden,
namely the choice of all six outermost vertices (shown in the bottom
snowflake of the figure), because this would
leave the central vertex undominated. The other 95 possibilities lead
to valid \mds s.
Thus the star of $k$ snowflakes has at least
$95^k$ \mds s, as claimed, and the growth constant $\lambda$ cannot be
smaller than
$\lim_{k\to\infty} (95^k)^{1/(13k+2)} = \sqrt[13]{95}$.
We have ignored the \mds s that don't contain~$a$, but their number is negligible: it is
$64^k$.


A tree that approaches the upper bound more tightly is obtained
by omitting
 the vertex $b$, but it is not so
straightforward to analyze.  Such a tree has $13k+1$ vertices and
$95^k-63^k+64^k+k\cdot32^{k-1}\ge 95^k$ \mds s. 
 Let us at least confirm the
leading term: The $95^k$ sets are the same ones as before. If we
subtract the $63^k$ cases where \emph{every} star has a neighbor or a
distance-2 neighbor of $a$ in $D$, we are sure that the vertex $a\in D$
can choose a private neighbor. 
 This establishes
the lower bound $95^k-63^k=95^k(1-o(1))$ on the asymptotic growth for
these trees.
The last two
terms of the formula are for the cases where $a\in D$ chooses itself as a private neighbor
or $a$ does not belong to $D$.

This family of trees gives asymptotically the largest
 number of \mds s that we know.
 It approaches
 the bound $\lambda^n$ with a multiplicative error that goes to
 $1/\lambda\approx 0.704$
as $k\to \infty$,
and this proves part 3 of Theorem~\ref{sizes}.
 We call these trees our \emph{record trees} and denote them by
$\mathrm{RT}_{13k+1}$.

We remark that, in the
original star of snowflakes,
 the $95^k$ \mds s containing the vertex $a$
are in fact \emph{minimum} dominating
sets: dominating sets of smallest size.
Since they are always a subset
of the \mds s,
the asymptotic growth constant $\lambda$ is valid also for 
 \emph{minimum dominating sets} in trees.

\section{Counting \MDS s of a Particular Tree: Dynamic Programming}
\label{counting}

\subsection{Combining rooted trees}

It is not difficult to compute the number of \mds s of a tree by
dynamic programming, and there are different ways to organize the computation.
For inductively building up a tree from smaller trees,
 it is
convenient
to mark an arbitrary vertex as the \emph{root} of the tree.
We  combine trees 
 with the following
\emph{composition} operation: We take two rooted trees $A$ and $B$ and
add an edge between the roots. The root of $A$ is kept as the root of
the result.
The basic building block for the construction is the singleton tree.
There are many ways in which
a given tree $T$ can be built up
 through a sequence of
compositions:
After selecting an arbitrary root vertex $r$
for $T$,
one picks an edge $rs$
incident to $r$ and removes it. This results in two trees, with roots
$r$ and $s$, from which the tree is composed. The two rooted trees are
further decomposed recursively.
In the following, 
we will specify a subtree by its vertex set $A\subseteq V$ and its root~$r$.

We want count \mds s bottom-up, following the composition.
 In this
process, we have to count \emph{partial solutions}, i.e., subsets
$D\subseteq A$
that have the potential to become a \mds\ when more components are connected
to the root $r$.
In Section~\ref{Preliminaries} we have characterized \mds s 
by requiring that every vertex is \emph{legal}.
The subtree $A$ is connected to the rest of
the tree by edges incident to $r$;
therefore, 
 $r$ itself need not be
legal
in a partial solution. 
Every vertex $a\ne r$, however, 
 must be legal: It
 is dominated, and if it belongs to $D$, then it
has a private neighbor.

\begin{figure*}[hbt]
  \centering
  \includegraphics[scale=0.97]{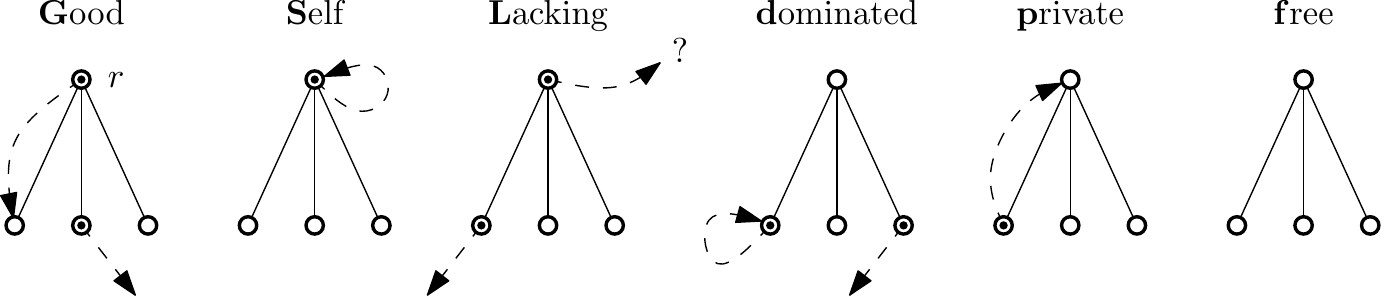}
    \Description{Graphical illustration of the six categories}
  \caption{Six types of partial solutions for a rooted tree.
We show the root $r$ and its neighbors in some typical configuration.
The vertices belonging to
    $D$ are marked. The dotted arrow indicates the private neighbor for a vertex.}
  \label{solutions}
\end{figure*}

\begin{table}[bt]
  \centering
$$
\vcenter{\hbox{%
    \includegraphics[scale=1.1]{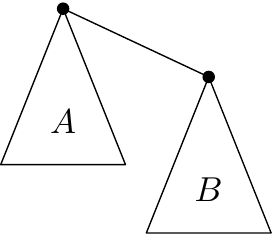}
  }}
\ifsoda
\else
\hskip 1,52cm
\fi
\vcenter{\hbox{\begin{tabular}{|cc|cccccc|}
\hline
&    & \multicolumn6{c|}{$B$}\\
&& \textbf{G}&\textbf{S}&\textbf{L}&\textbf{d}&\textbf{p}&\textbf{f} \\\hline
&\textbf{G} & \textbf{G}&$-$&$-$&\textbf{G}&$-$&\textbf{G} \\
&\textbf{S} & \textbf{L}&$-$&$-$&\textbf{S}&$-$&\textbf{G} \\
\smash
{\hbox{\lower 6pt\hbox{$A$}}}
&\textbf{L} & \textbf{L}&$-$&$-$&\textbf{L}&$-$&\textbf{G} \\
&\textbf{d} & \textbf{d}&\textbf{d}&$-$&\textbf{d}&\textbf{d}&$-$ \\
&\textbf{p} & $-$&$-$&$-$&\textbf{p}&\textbf{p}&$-$ \\
&\textbf{f} & \textbf{d}&\textbf{d}&\textbf{p}&\textbf{f}&\textbf{f}&$-$ \\
\hline
  \end{tabular}}}
$$
  \caption{The category 
 when a tree of type $B$ is attached
    as a child to a tree of type $A$. The symbol ``$-$'' indicates that the
    result is not valid.}
  \label{tab:combine}
\end{table}

\subsection{Combining partial solutions}

\ifARXIV
By sitting down and thinking how to compose partial solutions, one
will discover that six
\else
Six
\fi
types of partial solutions must be distinguished, see Figure~\ref{solutions}:
When the root belongs to $D$, there are three categories, which we denote with
capital letters:
\begin{itemize}
\item \textbf{G}ood. The root $r$ has a private neighbor among its neighbors.
\item \textbf{S}elf. The only private neighbor of the root $r$ is $r$ itself.
\item \textbf{L}acking. The root $r$ does not yet have a private neighbor.
The private neighbor needs to be found among the neighbors
that will still be attached to~$r$.
\end{itemize}
When the root is not part of $D$, there are three more categories, indicated by
small letters:
\begin{itemize}
\item \textbf{d}ominated. The root $r$ is dominated by some
  neighbor in $D$, and each vertex in $D$ has a private neighbor
  different from $r$.
\item \textbf{p}rivate. There is vertex 
in $D$ whose only private neighbor is the root.
\item \textbf{f}ree. The root has no neighbor in $D$.
A neighbor that will dominate $r$ needs to be found in the components
that will still be attached to~$r$. 
\end{itemize}

Table~\ref{tab:combine} shows the resulting category of a composite
tree depending on the category of the components.
Let us give an 
example:
When composing a partial solution of type $\mathbf{L}$ for a
 tree $A$ with root $r$
and a partial solution of type $\mathbf{f}$ for
a tree $B$, the
root $s$ of $B$ can be used as the private neighbor for $r$,
and at the same time, $s$ has found a dominating vertex, namely~$r$.
The result
will be of type \textbf{G}.
Some compositions are not valid:
For example, when
$B$ is of type $\mathbf{p}$,
the root $s$ of $B$ is the only
private neighbor of some vertex below it.
When this is combined with a tree
 $A$ of type $\mathbf{G}$,
 $\mathbf{S}$, or
 $\mathbf{L}$,
 $s$ can no longer function as a private neighbor,
 because it is adjacent to
 the root of $A$, which belongs to~$D$.
 The other entries of the table can be worked out similarly.

\subsection{Characteristic vectors}
\label{characteristic-vectors}

For a rooted tree, we
record the number of partial solutions of each type
in a 6-vector
$v=(G,S,L,d,p,f)$.
Table~\ref{tab:combine} can be directly translated into the formula for
the vector obtained by combining two subtrees $T_1$ and $T_2$
 (written as 
column vectors):
\begin{equation}
  \label{eq:circ}  
\begin{pmatrix}G_1\\S_1\\L_1\\d_1\\p_1\\f_1\end{pmatrix} \star
\begin{pmatrix}G_2\\S_2\\L_2\\d_2\\p_2\\f_2\end{pmatrix} :=
\begin{pmatrix}
G_1G_2+G_1d_2+G_1f_2+S_1f_2+L_1f_2\\
S_1d_2\\
S_1G_2+L_1G_2+L_1d_2\\
d_1G_2+d_1S_2+d_1d_2+d_1p_2+f_1G_2+f_1S_2\\
p_1d_2+p_1p_2+f_1L_2\\
f_1d_2+f_1p_2
\end{pmatrix}
\end{equation}
The \emph{final categories} are those
 partial solutions that can stand alone
as a minimal dominating set:
 \textbf{G}, \textbf{S}, \textbf{d}, and \textbf{p}.
Therefore, the total number $M(T)$ of minimal dominating sets of a tree $T$ with vector
$(G,S,L,d,p,f)$ is calculated by the linear function
\begin{equation}
  \label{eq:total}
 \bar M(G,S,L,d,p,f):=G+S+d+p. 
\end{equation}
A single-vertex tree has category \textbf{S} when the vertex belongs
to $D$, and category \textbf{f} if $D=\emptyset$.
Thus, a single-vertex tree has
 the vector
 \begin{equation}
   \label{eq:start}
   v_0 
   :=
   (0,1,0,0,0,1)  .
 \end{equation}
 This provides the starting condition
for the recursion.

\begin{figure}[htb]
  \centering
  \includegraphics{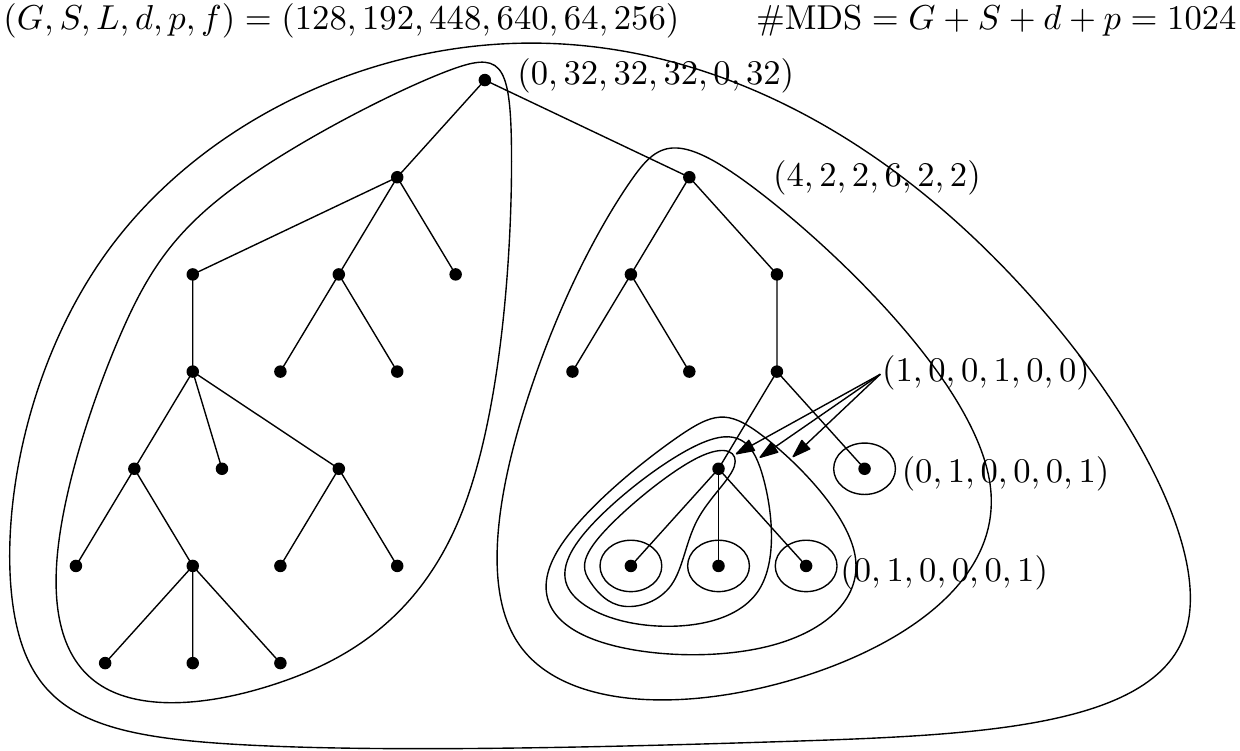}
  \Description{A worked example, showing the vectors for some selected
    subtrees and the whole tree}
  \caption{Calculating the number of \mds s of a tree bottom-up}
  \label{fig:example}
\end{figure}

We have now all ingredients for a straightforward counting algorithm
for the \mds s of a tree: choose a root, recursively decompose the
tree into smaller parts, compute the vectors for all parts in a
bottom-up way, and apply the operation $\bar M$
from~\eqref{eq:total}
to the result vector.
Figure~\ref{fig:example} shows a partially worked example.

\iffull
All the knowledge about 
the possible
number of \mds s that a tree with $n$ vertices can have is actually
embodied in these formulas: the starting vector \eqref{eq:start}, the
composition operation \eqref{eq:circ} in terms of the bilinear operation $\star$, and the terminal
formula~\eqref{eq:total}. 

Before we embark on studying these formulas
from a quantitative viewpoint, we will use them for designing an
enumeration algorithm.

\section{Listing all \MDS s of a Tree}
\label{listing}

\ifsoda
In Section~\ref{counting},
\else
In the previous section,
\fi
the composition rules in 
Table~\ref{tab:combine} have been used to design
a dynamic-programming algorithm for \emph{counting} \mds s,
based on
the recursion~\eqref{eq:circ} for the \emph{number} of
partial solutions of each category.
We can reinterpret~\eqref{eq:circ} as an implicit
representation of the
 \emph{set} of
partial solutions. 
For instance,
Table~\ref{tab:combine} tells us that
each solution of category \textbf{S} for a subtree $A$ and each solution of
category \textbf{G} for $B$, when taken together, give rise
to a solution of category \textbf{L} for the combined tree.
Accordingly,
we find the term $S_1G_2$ in~\eqref{eq:circ}, but we now
interpret the multiplication
as a sort of Cartesian product operation, combining all solutions of
one set with all solutions from another set. The $+$ operation is
interpreted as set union.

Below, we will first model the dynamic-programming recursion as a
 directed acyclic graph.  Based on this implicit
 representation
of the solutions, we will then develop an
output-sensitive algorithm for listing all solutions.


\subsection{The expression DAG}
\label{DAG}
%
The directed acyclic graph (DAG)
for representing all solutions in a tree $T$ has
 three kinds of nodes:
\emph{basis nodes},
\emph{product nodes}, and \emph{union nodes}. 
Each node $K$ is \emph{associated} to some subtree $A
$ of $G$ and it implicitly
\emph{represents} a some 
class $R(K)\subseteq 2^A$ of vertex subsets of $A$, namely the partial
solutions of a certain category.  

A \emph{basis node} $K$ has no outgoing arcs, and it is associated to
a singleton subtree $A=\{a\}$. 
Its role is to {declare} that the vertex $a$ is in $D$ or does not belong
to $D$.
Accordingly, it represents
the set $D=A=\{a\}$ itself ($R(K)=\{\{a\}\}$) or
the empty set ($R(K)=\{\emptyset\}$).
For uniformity, we also allow a basis node to
 represent no set ($R(K)=\{\}$), but we will eventually get rid of such nodes.
 
 A \emph{product node} $K$ has two outgoing arcs to neighbors $K_1$ and $K_2$
that are associated to disjoint subtrees $A_1$ and $A_2$. The product
node is then associated to $A_1\cup A_2$, and it represents the vertex
subsets obtained by combining each subset of $A_1$ represented by
$K_1$ with each subset of $A_2$ represented by $K_2$:
$$
R(K) = \{\, D_1\cup D_2 \mid D_1 \in R(K_1), D_2 \in R(K_2)\,\}
$$

A \emph{union node} $K$ has 
two outgoing arcs to neighbors $K_1,K_2$
that are 
associated to the same subtree~$A$. The union node is
then also associated to $A$, and it represents the \emph{disjoint union} of
its successor nodes:
$$
R(K) =  R(K_1)\cup R(K_2)
$$

One node of
the DAG is designated as the \emph{target node}
that represents the final solution set.
 It has no incoming
arcs, and it is
associated to the vertex set $V$ of the whole tree.
We draw the arcs from top to bottom, with the target node topmost and
the basis nodes at the bottom.

With these types of nodes, it is straightforward to build an
\emph{expression DAG} $\mathcal{X}
$ that represents the
\mds s of a tree $T$.
$\mathcal{X}$
 has
 a node for each subtree that occurs in the
composition sequence and for each category.
Additional nodes are necessary for intermediate results when forming
multiple unions.
Figure~\ref{fig:ex} illustrates the construction with an example of the node $(C,\mathbf{L})$ for a
rooted subtree $C$ that is composed of two subtrees $A$ and $B$.
This node 
represents all partial solution of category
\textbf{L} in the subtree~$C$.
\begin{figure}[htb]
  \centering
  \ifsoda
  \includegraphics[width=\columnwidth]{dag}
  \else
  \includegraphics{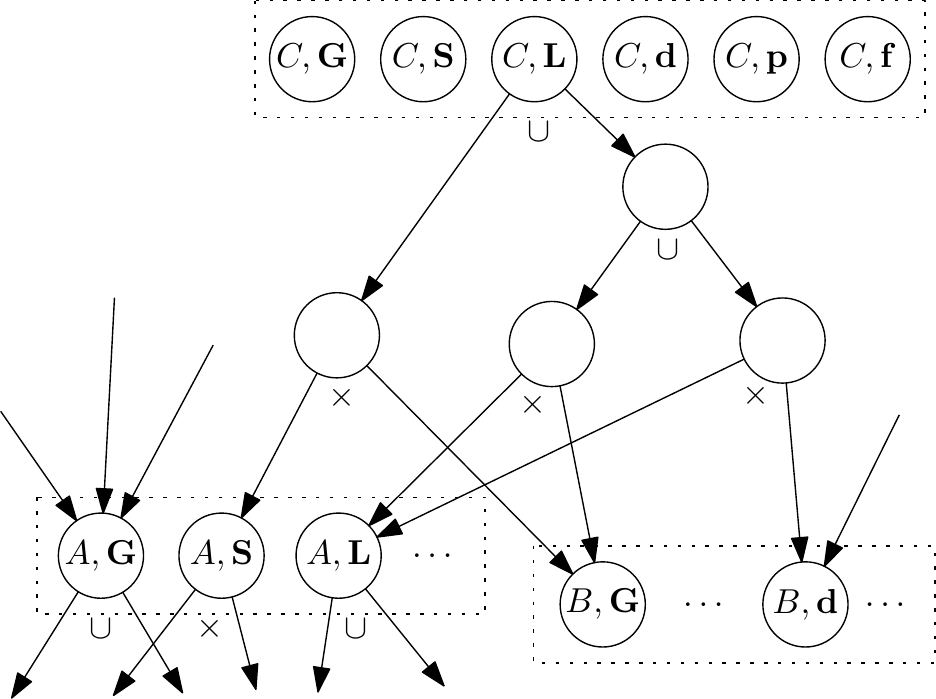}
\fi
  \caption{A part of the DAG $\mathcal{X}$ corresponding to the third entry
$S_1G_2+L_1G_2+L_1d_2$ in~\eqref{eq:circ}.
Union and product nodes are marked by $\cup$ and $\times$.}
  \label{fig:ex}
\end{figure}

The whole construction has $6n + 34(n-1)+3$ nodes.  $6$ nodes
are used to represent each singleton tree: One node represents the
singleton set $\{\{a\}\}$, of category \textbf{S}, another one
represents the empty set $\{\emptyset\}$, of category \textbf{f}, and
the four others represent no set.  There are $n-1$ composition steps,
one for each edge of $T$, and for each composition we need $34$ nodes:
$34=14+20$ is the number of additions and multiplications on the
right-hand side of \eqref{eq:circ}.  Finally, we need 3 union nodes to
compute the union of the categories \textbf{G}, \textbf{S},
\textbf{d}, and \textbf{p} for the whole tree, corresponding to the
total sum $\bar M=G+S+d+p$.  It is important to note that all union
nodes in this construction represent disjoint unions, as
every partial solution belongs to a unique category.
Another important property of the tree is that
a path can go through at most 8 consecutive union nodes:
\label{at-most-8}
The largest
number of additions for a single entry of \eqref{eq:circ} if 5; we
have to add
3 for evaluating $\bar M$.
The bound of 8 can be reduced to~4 if we care to balance the
network of union nodes.

We can reinterpret $\mathcal{X}$ 
as an
arithmetic circuit, by viewing union and product nodes as 
addition and multiplication gates,
and basis nodes as inputs with values 0 or 1.
Then the value computed in each node equals the number of subsets
represented by that node, and the computation modeled by this circuit
is nothing but our counting algorithm of Section~\ref{counting}.

\subsection{Pruning of nodes}
We now get rid of unnecessary nodes.
In a first sweep we proceed upward from the basis nodes towards the
target and eliminate all nodes representing the empty set. (They
correspond to the gates
that have value 0.) These are first of all the basis nodes of categories
\textbf{G},
\textbf{L},
\textbf{d}, and
\textbf{p}.
Continuing towards the target node, we eliminate all union nodes without successor, and all product nodes
that have lost at least one successor.

In a second, downward, sweep from the target towards basis nodes, we
delete
all nodes that do not contribute towards the result.
These are all nodes without predecessor, except for the target node.
In particular, intermediate results that would only be multiplied by 0 are
discarded.

In a final clean-up step,
we eliminate each union node $K$ with a single successor $K'$ and introduce
shortcut arcs from the predecessors of $K$ to $K'$.

Every node of the resulting DAG is now ``useful'': it represents a
nonempty set, and it is computed through a nontrivial operation from its children. When the DAG is viewed as an arithmetic circuit, it starts with
ones and performs multiplications and additions of positive numbers
that will eventually contribute to the total number of \mds s. Thus,
we need not worry about computing with excessively
big numbers while the eventual result is small.
For any tree $T$ of size $n$ we can evaluate the
number $M(T)$ with $O(n)$ additions and multiplications of numbers
that are bounded by $M(T)$, with $O(n)$ overhead.
(It is likely that even a straightforward application of the
composition rules \eqref{eq:circ} without pruning
never involves numbers that substantially exceed $M(T)$, but we have
not tried to show this.)

\subsection{The enumeration algorithm ENUM1\ifsoda.\fi}
\label{enumeration}
The idea of 
the algorithm
 is clear: to enumerate the solutions represented by a union
node, we have to enumerate solutions for the two successor nodes
in sequence.
For product nodes, 
the results of the
successor nodes must be
combined in all possible ways,
by cycling through them
in two nested loops.
The real ``work'' is done only in the basis nodes:
deciding whether a particular node belongs to the \mds\ $D$ or not.
We arbitrarily
order the two successors of union and product nodes, so that we can
speak of the first and second \emph{child}. (We use the term ``child''
although $\mathcal{X}$ 
is not a tree.)

The program is easiest to write
in a language like \textsc{Python} that supports
generator functions, 
see
Figure~\ref{python}.
\ifsoda
\begin{figure}[bt]
\else  
\begin{figure}[hbt]
\fi  
\centering
\ifjustsyntaxchecking\else
 \begin{minipage}{34em}
\begin{verbatim}
class Basis_node_S(Node):
   def enumerate_solutions(self):
      a = self.vertex
      yield [a]   # category S
\end{verbatim}
\begin{verbatim}
class Basis_node_f(Node):
   def enumerate_solutions(self):
      yield []    # category f, the only solution is the empty list
\end{verbatim}
\begin{verbatim}
class Union_node(Node):
   def enumerate_solutions(self):
      for D in self.child1.enumerate_solutions():
         yield D
      for D in self.child2.enumerate_solutions():
         yield D
\end{verbatim}
\begin{verbatim}
class Product_node(Node):
   def enumerate_solutions(self):
      for D1 in self.child1.enumerate_solutions():
         for D2 in self.child2.enumerate_solutions():
            yield D1+D2 # concatenation of lists D1 and D2
\end{verbatim}
\begin{verbatim}
# main call:
for D in target_node.enumerate_solutions():
   print D # or otherwise process D
\end{verbatim}
\end{minipage}
\fi

\caption{Recursive enumeration algorithm in \textsc{Python}}
\label{python}
\end{figure}
%
Each node of $\mathcal{X}$ is represented by a \textsc{Python} object.
The different node types are subclasses of a common
superclass \texttt{Node} whose definition is not shown.
What is also omitted is the code to generate the graph and to set
the \texttt{vertex} or
the \texttt{child1} and \texttt{child2}
attributes of the nodes.

The \texttt{yield} statement of \textsc{Python} suspends the execution of the current
function until the next generated element is requested in the
\texttt{for}-loop in which the function is called.
 Different 
generator functions and different nested loops are simultaneously
active, and they interact like coroutines.
The first parameter \texttt{self} of the functions is just
\textsc{Python}'s convention to refer to the object to which a method
is attached.

The \textsc{Python} library actually provides standard functions for achieving precisely
the effect of the enumeration procedures in the union and product
nodes: the functions \texttt{itertools.chain} and
\texttt{itertools.product} from the \texttt{itertools}
package.  For clarity, we
wrote the
loops explicitly
instead of using these functions.

As currently written in Figure~\ref{python}, the generation takes more than linear time per
solution, because each solution is built up by concatenating
shorter lists \texttt{D1} and \texttt{D2} into longer lists
\texttt{D1+D2}, which is not a constant-time operation in \textsc{Python}. This has been done to make
the program clear, but it is easy to fix:
We can either use linked lists, or we just let each basis node set or
clear
a bit in a bit-vector representation of the solution.
In the last variant, the program for a basis node of category
\textbf{S} would be as follows:
\medskip\par\indent
\ifjustsyntaxchecking\else
 \begin{minipage}{34em}
\begin{verbatim}
      i = self.vertex_number
      D[i] = True  # category S
      yield None
\end{verbatim}
 \end{minipage}
\medskip\par\noindent
\fi
and accordingly with \texttt{False} for category~\textbf{f}.
The solution is maintained in the global variable \texttt{D},
which is a list
of Boolean values.
No partial solutions are ever returned to the calling subroutine, and
the combination of the solutions can be bypassed.
 All \texttt{yield} statements of the program are changed so that they just produce the dummy element
 \texttt{None}.
 We will refer to this version as algorithm ENUM1.
If desired, the solution can be constructed in any suitable form at the target node from the
bit vector \texttt{D} in linear time.

The enumeration works as follows:
When a new solution is needed, a call
\verb|enumerate_solutions|
is initiated at the target node and proceeds towards the basis nodes. For a union node, one child is
entered,
and for a product node, the algorithm enters both children or only the 
second child, in case we are in the inner loop and the solution
\texttt{D1} of the first child remains fixed. Eventually, at most
one basis node is entered for each vertex, and there it is decided whether this
vertex belongs
to the solution $D$ or not.
The visited nodes form a subtree of $\mathcal{X}$ with at most $n$ leaves.
As we have observed, there can be at most 8 consecutive levels of
union nodes where the tree
does not branch.
From this, one can conclude that
the subtree of visited nodes has linear size.

However, 
there is a subtlety
in the way how  generators are handled in
\textsc{Python},
which makes this argument invalid:
When a loop like
\begin{equation*}
  \hskip\parindent\vbox{\noindent
  \texttt{for $x$ in
  }$\langle\textit{generator-function}\rangle$\texttt{: }
  \dots}
\end{equation*}
loops over $k$ successive elements $x$, the
\textit{generator-function} is actually called $k+1$ times. In the $(k+1)$\nobreakdash-st
iteration,
it will raise the
\texttt{StopIteration} exception to signal that there are no more
items.
Thus, in a union node, for example, the algorithm does not always descend into
just \emph{one} of the two children in the clean way as we supposed in our
description. It might call
  \verb|self.child1.enumerate_solutions()|,
 only to receive a
\texttt{StopIteration} exception and subsequently call
\verb|self.child2.enumerate_solutions()|.

Despite this behavior,
the runtime between successive solutions
is still $O(n)$. This fact requires a more elaborate analysis, which we will give
in Section~\ref{python-compare}.
Here
it is 
important that the number $k$ of elements generated by every generator
function is positive, due to the preparatory pruning of the
expression DAG.
Before that,
in Section~\ref{enum2},
we will
 describe and analyze a different process, ENUM2,
 for which the above argument goes through in a clean way.
 The analysis of ENUM1 in Section~\ref{python-compare} builds on these results. 
%
In the next section, we will first discuss a possibility for  optimizating the
\emph{total} generation time. 

\subsection{Optimizing the overall runtime by reordering the children}
\label{sec:optimize}

As we have argued, and as we will show in
Section~\ref{python-compare}, the algorithm takes $O(n)$ time per solution.  In a setting where we want to examine each
solution explicitly, this is optimal and leaves no room for
improvement (at least if the typical solutions $D$ are not much
smaller than~$n$).

Algorithm ENUM1 does not treat the children of a
product node equally: While the solutions for child 1 are only
enumerated once, the solutions for child~2 are enumerated again and
again as part of the inner loop.  
One may try to optimize the
running time by choosing the best order.
Potentially, one may even achieve
sublinear average time per solution.
 
In fact, in most enumeration tasks, an explicit list that can
be stored is not what is actually needed, but one wants to run
through all solutions, for example with the objective to evaluate them and
choose the best one.  Often, such an evaluation can be
maintained incrementally: It is cheaper to \emph{update} the objective
function of $D$ when a vertex is inserted or deleted instead of
computing it from scratch.
In such a setting, if makes sense to strive for sublinear average time.
 Since the basic operation of our enumeration algorithm is the
 insertion or deletion of single elements, the runtime of Algorithm
 ENUM1 gives an appropriate model for such an application case.

 Let us therefore analyze the runtime for some product node $K$.
 Assume that child~$i$ represents $C_i$ solutions, and
 $t_i$ is the average time per
 solution, i.\,e.,
it takes time
 $t_iC_i$ to enumerated all solutions.
  Then, up to constant factors, the total time for node~$K$
 is
 \begin{displaymath}
   C_1C_2 + C_1t_1 + C_1C_2t_2.
 \end{displaymath}
 Here, the first term
 $C_1C_2$ acounts for the time spent internally in the enumeration
 procedure for node~$K$
 (putting together
 the solutions, passing them to the parent node, etc.), without the
 recursive calls.
 For this analysis, the extra \texttt{StopIteration} call at the end of the loop
 does not hurt us, because it would
only  change
 $C_1C_2$
 to $C_1C_2+1$, and thus, for the overall runtime, it would influence only the constant factor.

 The resulting average time per solution is
 \begin{displaymath}
   t = 1 + t_1/C_2 + t_2
   .
 \end{displaymath}
This has to be compared against
 \begin{math}
   t' = 1 + t_1 + t_2/C_1
 \end{math}.
The typical case is when the numbers $C_i$ are large; then the
term that is divided by $C_i$ becomes negligible, and the optimal
choice gives
\begin{equation} \label{combine-trees}
  t \approx 1+ \min\{t_1,t_2\}.
\end{equation}

For a union node, we have total time of 
 \begin{displaymath}
   C_1+C_2 + C_1t_1 + C_2t_2
=    C_1(t_1+1) + C_2(t_2+1).
 \end{displaymath}
Thus, a union node effectively adds a constant overhead to each
solution.
One can optimize the structure of a tree of union nodes into a
Huffman tree. However,
since the number of consecutive levels
 of union nodes is already bounded by 8, this will change the runtime at most
 by a constant factor. 

 For a given expression DAG,
it is straightforward to compute the required quantities bottom-up and to reorder the
children appropriately.
Moreover, a given tree $T$ has many recursive decompositions into
subtrees,
and it might be interesting to choose a best one.
 Formula~\eqref{combine-trees} suggests that the
runtime should depend on the shortest path from the root to a leaf (basis node).
More precisely, such a short path
should exist from every product node
that is reachable from the target node through a sequence of union
nodes.
On the other hand, a short path to a leaf indicates a small subtree,
and for small subtrees, the assumption
under which the approximate formula \eqref{combine-trees} was derived, namely
that the number of solutions is large, is not satisfied.
We leave it as an open problem to find the right balance and to analyze
 the speedup that can be achieved in general with these ideas.


 \begin{figure}[htb]
  \centering
  \includegraphics{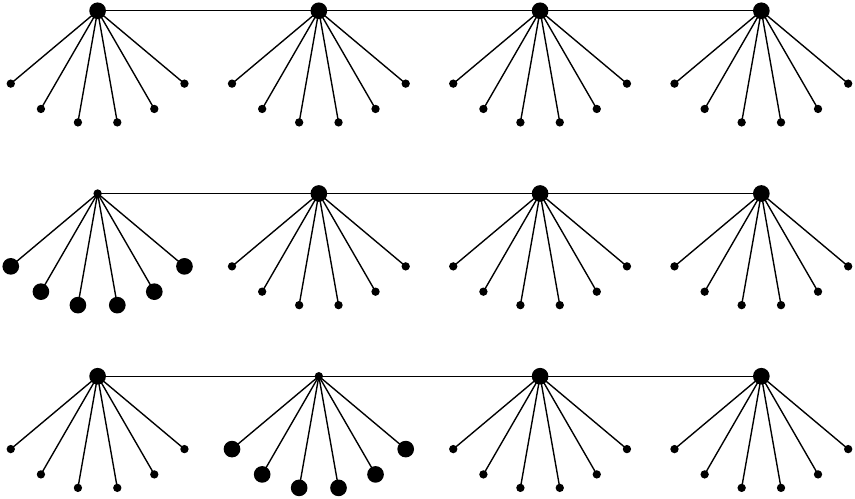}
\Description{A graph with a chain of four stars, and three examples of
  \mds s. The leaves that are
  adjacent to a common vertex must either all belong to $D$ or not.}
  \caption{Minimal dominating sets in a chain of stars}
  \label{fig:stars}
\end{figure}

However, there is
a limit on the speedup that one can hope for:
The tree in
Figure~\ref{fig:stars}
consists of many clusters of leaves that are adjacent to a common
vertex like in a star. By
 Observation~\ref{obse}(\ref{twins}), all these twins must belong to a
 \mds\ together. Thus, to go from one \mds\ to another, one has to
 completely swap at least one such cluster into or out of the solution.
 With $k$ stars of size $n/k$, there are $2^k$ solutions,
 and it takes at least $n/k$ time just to swap nodes in and
 out of any solution. Taking $k\approx a \log_2 n$ for some constant $a$,
 this gives an
 example with
$\Theta(n^a)$ solutions and a total
 running
 time
 $\Omega(n^a\times n/\log n)$.
 This rules out a speed-up by more than a logarithmic factor.

In view of this example, 
it makes sense to lump
clusters of twins together as a preprocessing step.
From each cluster of twin leaves, one representative is chosen,
and the other vertices go along with that representative.
Essentially, this means that we delete all leaves except one representative from
each cluster, or
in other
words, we consider only graphs without twins.

It seems that such graphs always have an exponential number of \mds s.
We found empirically that, for $2\le n\le 70$,
 the number of solutions is
 at least 
$2^{n/3}$. We calculated this by adapting
 the algorithm from Section~\ref{upper} 
below to the \emph{minimization} of the number of solutions.
It turned out that when
$n$ is of the form $3k-1$, the tree without twins that has the
smallest number of \mds s is the
extended comb with $k$ teeth shown in
Figure~\ref{fig:comb}c.
From each of the $k$ teeth, one can independently choose one of the
two vertices. Such a selection can be completed into a unique \mds\ by adding
an appropriate subset of the $k-1$ intermediate vertices between the
teeth;
thus, there are exactly $2^k=2^{(n+1)/3}$ \mds s in this example.
For $n=3k-2$, one can get a tree with the same number $2^k$ solutions by removing the leftmost or rightmost leaf of
Figure~\ref{fig:comb}c.
For $n=3k\ge 6$, the best tree has $\frac 74 \cdot 2^k$ solutions.
These statements are not proved to hold in general. The proof technique of
Section~\ref{upper2} should be applicable,
but we did not try.

The exponential number of solutions for trees without twins gives hope
that one might be able to enumerate the \mds s in substantially sublinear
average time, because occasional expensive updates can be
amortized over a large number of outputs.

\subsection{Implementation by message passing: Algorithm ENUM2}
\label{enum2}
We give now a more explicit description of the enumeration procedure
as a message-passing algorithm, without relying on
the generator framework.
 At any time, there is one
active node of the DAG.
This node sends a message to one of its neighbors, and the action passes
to that neighbor.
The nodes maintain private state variables.

There are two types of \emph{request messages}, which always flow downward in the network:
VISIT and V+NEXT.
There are two types of \emph{reply messages}, which
 flow upward in response to the request messages:
DONE and LAST.

\begin{figure}[t]
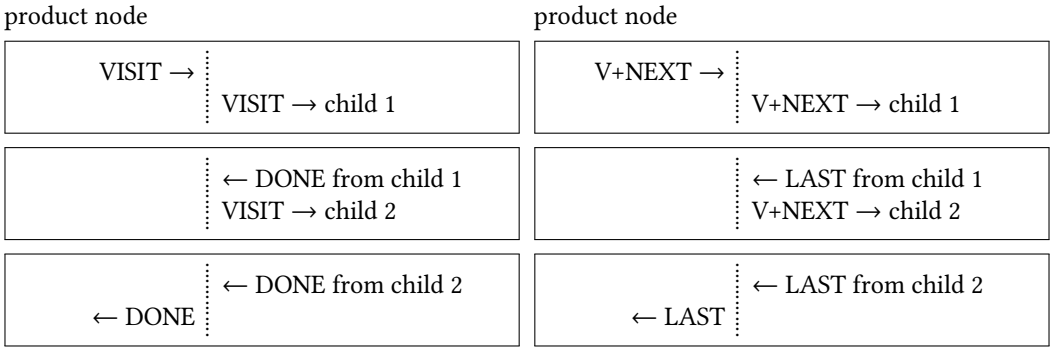

  \centering

\hbox to \hsize{%
\commandbox{product node}{%
  VISIT $\to$&\\
& VISIT $\to$ child 1\\
}
\hfill
\commandbox{product node}{%
  V+NEXT $\to$&\\
& V+NEXT $\to$ child 1\\
}}

\hbox to \hsize{%
\commandbox{}{
& $\from$ DONE from child 1\\
& VISIT $\to$ child 2\\
}
\hfill

\commandbox{}{
& $\from$ LAST from child 1\\
& V+NEXT $\to$ child 2\\
}}

\hbox to \hsize{%
\commandbox{}{
& $\from$ DONE from child 2\\
$\from$ DONE\\
}
\hfill
\commandbox{}{
& $\from$ LAST from child 2\\
$\from$ LAST&\\
}}
  \caption{Program for a product node}
  \label{fig:product}
\end{figure}

\begin{figure}[t]
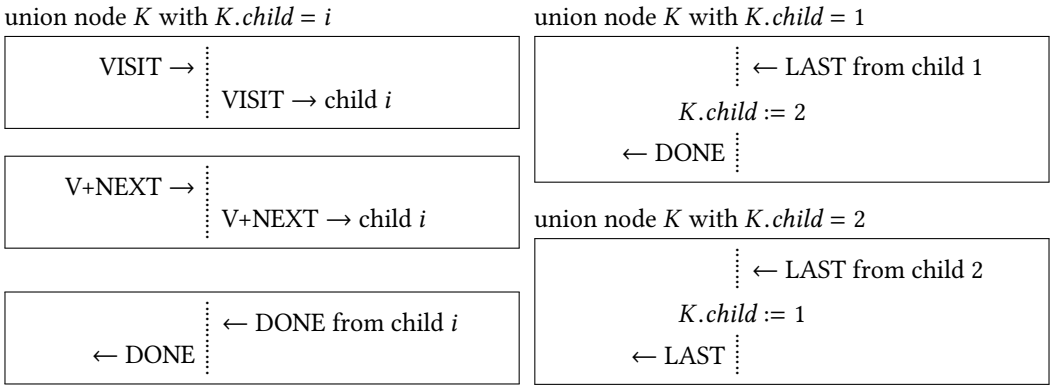

  \centering

\hbox to \hsize{%
\vtop{
\hrule height 0pt
\commandbox{union node $K$ with $K.\child=i$}{%
  VISIT $\to$&\\
& VISIT $\to$ child $i$\\
}
\vskip 5pt

\commandbox{}{
  V+NEXT $\to$&\\
& V+NEXT $\to$ child $i$\\
}

\vskip 11pt

\commandbox{}{
&$\from$ DONE from child $i$\\
$\from$ DONE\\}%
}

\hfill

\vtop{\hrule height 0pt
\hbox{union node $K$ with $K.\child=1$}
\vskip 3pt
\hbox{\fbox{%
\vbox{\hbox{\commandb{&$\from$ LAST from child 1\\}}
\moveright 2,6cm\hbox to 0pt{\hskip 1mm\hss $K.\child := 2$\hss}
\vskip 3pt
\hbox{\commandb{$\from$ DONE\\}
}}}}

\vskip 8pt

\hbox{union node $K$ with $K.\child=2$}
\vskip 3pt
\hbox{\fbox{%
\vbox{\hbox{\commandb{&$\from$ LAST from child 2\\}}
\moveright 2,6cm\hbox to 0pt{\hskip 1mm\hss $K.\child := 1$\hss}
\vskip 3pt
\hbox{\commandb{$\from$ LAST\\}
}}}}}}

  \caption{Program for a union node}
  \label{fig:union}
\end{figure}

\begin{figure}[t]
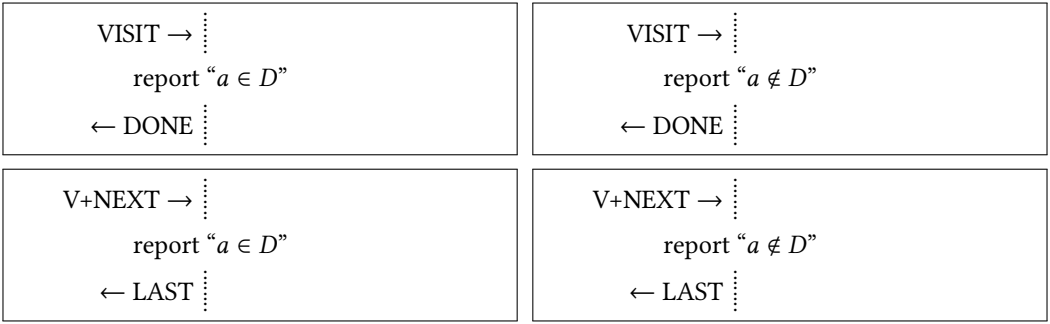

  \centering
\hbox to \hsize{%
\vbox{\hbox{basis node $K$ for vertex $a$, representing $\{\{a\}\}$}
\vskip 3pt
\hbox{\fbox{%
\vbox{\hbox{\commandb{VISIT $\to$&\\}}
\moveright 2,6cm\hbox to 0pt{\hskip 1mm\hss report ``$a\in D$''\hss}
\vskip 3pt
\hbox{\commandb{$\from$ DONE\\}
}}}}}
\hfil
\vbox{\hbox{basis node $K$ for vertex $a$, representing $\{\emptyset\}$}
\vskip 3pt
\hbox{\fbox{%
\vbox{\hbox{\commandb{VISIT $\to$&\\}}
\moveright 2,6cm\hbox to 0pt{\hskip 1mm\hss report ``$a\notin D$''\hss}
\vskip 3pt
\hbox{\commandb{$\from$ DONE\\}
}}}}}}
\vskip 1pt

\hbox to \hsize{%
\vbox{
\vskip 3pt
\hbox{\fbox{%
\vbox{\hbox{\commandb{V+NEXT $\to$&\\}}
\moveright 2,6cm\hbox to 0pt{\hskip 1mm\hss report ``$a\in D$''\hss}
\vskip 3pt
\hbox{\commandb{$\from$ LAST\\}
}}}}}
\hfil
\vbox{
\vskip 3pt
\hbox{\fbox{%
\vbox{\hbox{\commandb{V+NEXT $\to$&\\}}
\moveright 2,6cm\hbox to 0pt{\hskip 1mm\hss report ``$a\notin D$''\hss}
\vskip 3pt
\hbox{\commandb{$\from$ LAST\\}
}}}}}}

  \caption{Program for a basis node}
  \label{fig:basis}
\end{figure}

\begin{figure}[tb]
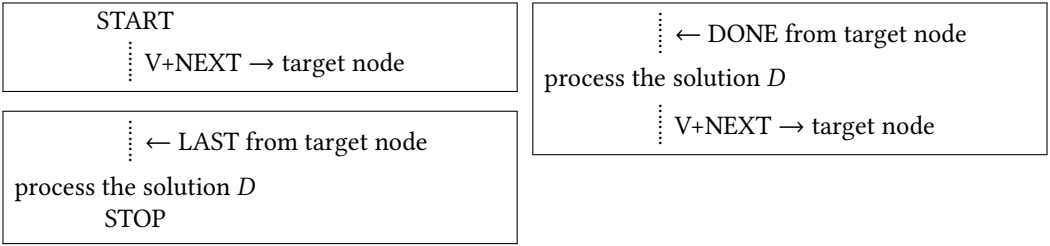
 

\newcommand{\commandbx}[1]{%
\valign{##\cr
\vskip 3pt
\hbox{\begin{tabular}{rl}
\omit\hskip 1,6cm\null&\omit\hskip 5,cm\\
#1
\end{tabular}}%
\vskip 1pt
\cr
\noalign{\kern -0,5mm\hskip - 5,cm}
\cleaders\vbox to 2pt{\vss\hbox{\small.}\vss}\vfill\cr
}\hskip 5,cm
}

  \centering

\hbox to \hsize{%
\vtop{\hrule height 0pt
\hbox{master node}
\vskip 3pt
\hbox{\fbox{%
\vbox{\moveright 1,6cm\hbox to 0pt{\hskip 1mm\hss START\hss}
\vskip 3pt
\hbox{\commandbx{&V+NEXT $\to$ target node\\}
}}}}

\vskip 6pt
\hbox{\fbox{%
\vbox{\hbox{\commandbx{&$\from$ LAST from target node\\}}
\moveright 1,6cm\hbox to 0pt{\hskip 1mm\hss process the solution $D$\hss}
\moveright 1,6cm\hbox to 0pt{\hskip 1mm\hss STOP\hss}
\vskip 3pt
}}}}
\hfil

\vtop{\hrule height 0pt
\hbox{master node}
\vskip 3pt
\hbox{\fbox{%
\vbox{\hbox{\commandbx{&$\from$ DONE from target node\\}}
\moveright 1,6cm\hbox to 0pt{\hskip 1mm\hss process the solution $D$\hss}
\vskip 3pt
\hbox{\commandbx{&V+NEXT $\to$ target node\\}
}}}}}}

  \caption{Program for the master node}
  \label{fig:master}
\end{figure}


The interaction follows a structured protocol: When a node $K$ sends a message to
one of its children $K'$ for the first time, a bidirectional \emph{channel}
between $K$ and $K'$ is established, and $K$ becomes the \emph{parent}
of $K'$, for the time being. Over this channel, 
the flow of messages is a strict alternation between downward requests
and upward replies:
%
%
\begin{equation}
  \label{eq:protocol}
\vcenter{\begin{quote}
  $\to$ V+NEXT\\
  $\from$ DONE\\
  $\to$ V+NEXT\\
  $\from$ DONE\\
  \dots\\
  $\to$ V+NEXT\\
  $\from$ LAST
\end{quote}}
\end{equation}
The meaning of this exchange is as follows:
V+NEXT stands for ``VISIT and ADVANCE TO NEXT SOLUTION''. It instructs the child
node to ``visit'' one solution, and to advance the internal variables
in the nodes of the DAG
so that the next visit will produce the next solution.
Successful
completion is signaled by the DONE message. The LAST message
signals in addition that the enumeration is completed and no more
additional solutions are available.
The state variables are then reset in such a way that the enumeration
will resume with the first solution when called again.
The node $K$ is then no longer the parent of $K'$,
and $K'$ is ready to another V+NEXT from a new parent.
If $K'$ represents $m$ solutions, this dialogue will finish after $2m$ messages.

The above dialogue can be interspersed with any number of VISIT/DONE
pairs of the following type:
\begin{equation}
  \label{eq:variation}
\vcenter{\begin{quote}
  $\to$ VISIT\\
  $\from$ DONE
\end{quote}}
\end{equation}
This will just visit the current solution but not advance the
pointers, so that the next VISIT or V+NEXT request will revisit the
same solution.

To record the current status of the enumeration,
every union node $K$ has an attribute $K.\child$ which is either 1 or 2.
At the beginning, all \child\ attributes are initialized to~1.
These are the only pointers that need to be explicitly
maintained. 
A union node $K$ will have an open channel to at most one of its
children at a time,
as selected by $K.$\child. A product node opens channels
to both children simultaneously.

We present the program in
Figures~\ref{fig:product}--\ref{fig:master}
in terms of simple \emph{patterns}:
For each node type and for each message that it potentially receives,
there is one pattern. The pattern prescribes some actions or some
variable change, and it terminates
with sending a message.
The message exchange with the parent is written on
the left of the dotted line,
the exchange with the
children occurs on the right side.
For example, the first box of code in
Figure~\ref{fig:product} says: If a product node receives a
VISIT request (from its parent), it sends a VISIT request to its first child.

We add a \emph{master node} with a single outgoing arc leading to the target
node (Figure~\ref{fig:master}). Its only job is to send V+NEXT
requests until the solutions are exhausted.

The program is very simple, but it is not immediate obvious 
from the patterns why it works.
To gain some understanding, we will first analyze the set of nodes that are visited when generating
one solution.

A subgraph $E$ of the expression DAG is called
a
\emph{\cwstree} if it contains
both children of every product node
in~$E$
and exactly one child of every union node
in~$E$.
%
The following lemma states some good properties of these graphs,
justifying their name ``\emph{well-structured} enumeration
\emph{trees}''.

\goodbreak

\begin{lemma}
  \label{structured-trees}
  \begin{enumerate}
  \item A \wstree\ is 
    a rooted directed tree, and its leaves
    are basis nodes.
  \item
    If the root of a
    \cwstree\ 
    is associated to the vertex set~$A$, then
   its leaves are in 
    one-to-one correspondence with the vertices of~$A$,
 \item A \cwstree\ 
   contains $\Theta(|A|)$ nodes in total.
  \end{enumerate}
\end{lemma}
\begin{proof}
  (1)
  By definition, a
 \wstree\ $E$ can
  branch only at product nodes.
Since the two children of such a node are associated to
disjoint subtrees of $V$, the two branches cannot
meet, and therefore $E$ is a tree.
(This justifies 
the terminology
of children
and parents that we are using.)
By definition,
the leaves of the tree can only be basis nodes.

(2)
This follows from the properties of the expression DAG:
When the tree branches at a product node, the associated set $A\subseteq V$ is split,
and at a union node, which has only one child, the
associated set is preserved.

(3) By (2), the tree has $|A|$ leaves.
As was argued
towards the end of Section~\ref{DAG}
on p.~\pageref{at-most-8}, a chain of
non-branching union nodes
has length at most~$8$.
It follows that the tree has $\Theta(|A|)$ nodes.
%
\end{proof}

We apply this lemma to bound the number of nodes visited by the algorithm:
\begin{lemma}
  Let $K$ be a node that is associated to a subtree $A$.
  We consider the period from the time when $K$ receives a message
  from its parent to the first time when it returns a message to its parent.
  \begin{enumerate}
  \item If $K$ receives a \textup{VISIT} message, the visited nodes
    form a \cwstree\ with root~$K$.  This tree is
 traversed in depth-first order.  No variables are changed, and
    the node will return a\/ \textup{DONE} message to its parent after
    visiting $\Theta(|A|)$ nodes.

  \item Consequently, if the node $K$ repeatedly receives
    \textup{VISIT} messages, the algorithm will revisit the same
    sequence of nodes again.
 
  \item If $K$ receives a {\textup{V+NEXT}} message, the algorithm
    will visit the same sequence of nodes as if a \textup{VISIT}
    message had been received. However, some variables may be changed,
    and the node may return a \textup{DONE} or a \textup{LAST} message
    to its parent.
  \end{enumerate}
\end{lemma}

\begin{figure}
  \centering

\hbox to \hsize{%
\vtop{\hrule height 0pt
\commandbox{union node $K$ with $K.\child=i$\strut}{%
  VISIT $\to$&\\
& VISIT $\to$ child $i$\\
&$\from$ DONE from child $i$\\
$\from$ DONE\\}}

\hfil
\vtop{\hrule height 0pt
\commandbox{product node\strut}{%
  VISIT $\to$&\\
& VISIT $\to$ child 1\\
&$\from$ DONE from child 1\\
& VISIT $\to$ child 2\\
&$\from$ DONE from child 2\\
$\from$ DONE\\}}}

  \caption{The VISIT operation
 from the viewpoint of a union and a product node}
  \label{fig:flow-VISIT}
\end{figure}


\begin{proof}
  (1) It is easy to check that a VISIT message leads only to VISIT and
  DONE messages.
The
union and product nodes behave as shown in Figure~\ref{fig:flow-VISIT}.
For a union node, the program goes to exactly one of the children, and for
a product node, it recursively visits each child.
The statement follows from
Lemma~\ref{structured-trees}.

(2) is an immediate consequence of~(1).

(3)
One can easily check this by looking at
the programs.  The only difference to a VISIT is that
some DONE
replies may be changed to LAST,
and 
the \child\ attribute of some union nodes may change.
\end{proof}
If we apply the lemma to the target node, this shows that Algorithm ENUM2 has only a linear delay between successive solutions.



\begin{figure}
  \centering
















\hbox to \hsize{\hskip -3mm
\valign{\vfil#\vfil\cr
\hbox{\includegraphics[scale=0.92]{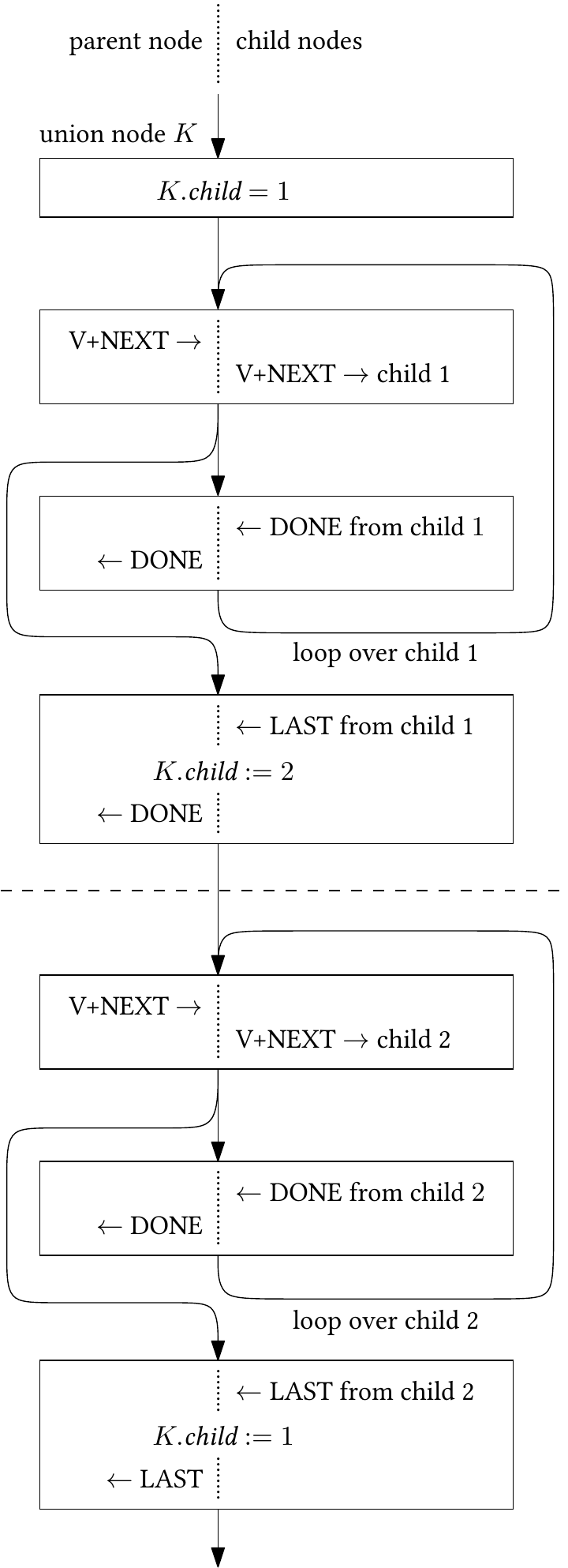}}\cr
\noalign{\hfil}
\hbox{\includegraphics[page=2,scale=0.92]{flowcharts-acm.pdf}}\cr%
}\hskip -3mm}
  \caption{Correctness is seen by observing the message flow from the viewpoint of a union node $K$ (left)
and
 from the viewpoint of a product node (right)
}
  \label{fig:flow-union}
\end{figure}

\subsection{Correctness}
To understand why the program is correct, let us focus on
the messages sent and received from a single node.
We prove by induction that
every node, when receiving a sequence of V+NEXT messages from
a parent, will follow
the protocol
\thetag{\ref{eq:protocol}}: 
Before each reply to the parent,
it will set up a solution in its associated subtree, and it will cycle
through all solutions and send back a LAST reply when it is done.

This is obvious for the basis nodes. For the union or product nodes,
we assume inductively that each child follows
the established protocol
\thetag{\ref{eq:protocol}}
from the first V+NEXT request to the LAST reply, and we get the
program flow
in
Figure~\ref{fig:flow-union}.
It is a matter of comparing the charts with the programs of
Figures~\ref{fig:product} and \ref{fig:union}
to check that they represent the true flow of actions.
The left part of
Figure~\ref{fig:flow-union} shows the process from the point of view
of a union node $K$. We clearly see the two successive
loops over the results of the two children. When the process
terminates, $K.\child$ is reset to~1. In this way, the node is reinitialized for the
next loop.
The right part shows a product node, and we see the loop over child~1
nested within the loop over child~2.
In both cases, the results are reported back to the parent in a cycle
ending with a LAST message. Thus the program is indeed a low-level
implementation of the loop structures for the recursive enumeration as
in the program ENUM1 of Figure~\ref{python} (except that for product
nodes, the nesting order of the two loops is different).

We have thus shown that the algorithm correctly generates all
solutions, with a linear delay between consecutive solutions.
The expression DAG
in the preprocessing phase can be constructed also in linear time,
thus
establishing
Theorem~\ref{theorem-enumeration}:
The \mds s of
  a tree with $n$ vertices can be enumerated with $O(n)$ setup time
and with $O(n)$ delay between successive solutions.

We give a few implementation hints that are not expressed in the
programs above.
A node must remember the parent from which it is currently receiving commands.
Alternatively,
the list of nodes that are still expecting replies can be maintained
as a stack. In this way, the parent node can simply be popped from the
stack when sending a message to it.
Besides this stack,
it may be convenient to maintain a \child\ attribute also for a
product node, in order to know from which child a message is received.

%


  






\subsection{Analysis of the 
  {Python} implementation ENUM1}
\label{python-compare}
As mentioned,
the concept of generator expressions in \textsc{Python}
uses a different convention for signaling the end of the data
stream.
Compared to Algorithm ENUM2, which signals the end of the data
simultaneously with the delivery of the last item, \textsc{Python}
does this only in response to the subsequent request, just like an
end-of-file condition is conventionally handled.
Such a behavior
is necessary in order to accommodate zero-length loops.
Here is a side-by-side comparison between the two conventions.

\smallskip
\indent
\begin{minipage}[t]{13em}
 Algorithm ENUM2 \eqref{eq:protocol}:
\begin{quote}
  $\to$ V+NEXT\\
  $\from$ DONE\\
  $\to$ V+NEXT\\
  $\from$ DONE\\
  \dots\\
  $\to$ V+NEXT\\
\lower 4,5pt\rlap{\hskip-1,2em\leaders\hbox{-\,}\hskip 8cm}%
  $\from$ LAST
\end{quote}
\end{minipage}
\
\begin{minipage}[t]{13em}
the \textsc{Python} convention: 
\begin{quote}
  $\to$ NEXT\\
  $\from$ DONE\\
  $\to$ NEXT\\
  $\from$ DONE\\
  \dots\\
  $\to$ NEXT\\
  $\from$ DONE\\
  $\to$ NEXT\\
  $\from$ STOP
\end{quote}
\end{minipage}

\bigskip

The NEXT message corresponds to \textsc{Python}'s \texttt{next()}
method,
and the STOP message is \textsc{Python}'s \texttt{StopIteration}
exception, 
which returns without producing a result.
After receiving a STOP message, a node might have to go again
to one of its children to produce an actual solution.
Therefore,
we need a more elaborate argument to show that the procedure still has only
linear delay.

We remark that the simpler protocol
 \eqref{eq:protocol} in the left column is only
possible because there are no null nodes that produce no
solution. Without this assumption, the linear-delay argument for the 
\textsc{Python} version ENUM1
that we are going to present would also break down.

\begin{figure}  \centering
\hbox to \hsize{
\valign{#\vfil\cr
\hbox{\includegraphics[scale=0.97]{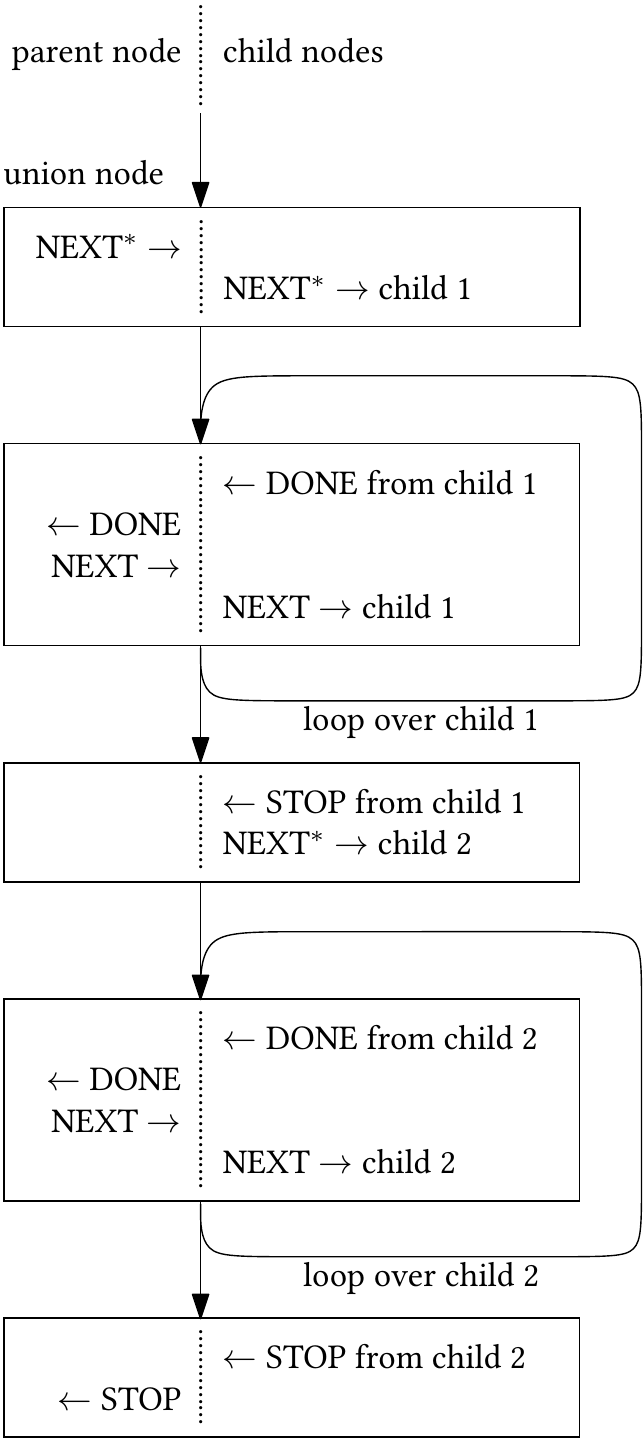}}\cr
\noalign{\hfil}
\hbox{\includegraphics[page=2,scale=0.972]{flowcharts-python.pdf}}\cr%
}\hskip -2mm
}
  \caption{Message flow of Algorithm ENUM1 in a union node $K$ (left)
and a product node (right)}
  \label{fig:flow-python}
\end{figure}

In Algorithm ENUM1, the union and product nodes do not perform any
operations except coordinating the loops over their children.  The
control flow inside a node that results from these
loops 
is shown in Figure~\ref{fig:flow-python}.  One difference to Algorithm
ENUM2 is that ENUM1 does not visit a basis node for each vertex in
every iteration.  In the inner loop of a product node, the solution of
the outer loop remains unchanged, and therefore it is not necessary to
enter the corresponding part of the tree.  This is the reason why
there is no need for a separate VISIT message like in Algorithm ENUM2,
(as opposed to V+NEXT).  The loops are terminated by STOP messages.
In the flow graphs of Figure~\ref{fig:flow-python} the very first NEXT
message that starts an iteration has been marked with a star. This is
when the node is entered by calling the function
\verb|enumerate_solutions|, while subsequent NEXT messages correspond
to the cases when the node is re-entered after a \emph{yield}
statement.

A \emph{visit} of a node is the time between receiving a request from
a parent and sending back a reply,
including recursive visits of descendent nodes.
When a node replies DONE after ``producing'' a valid solution, we call
this a
\emph{proper visit}.
When a node replies STOP to signal that there are no more solutions, we
speak of a
\emph{dummy visit}.
When a node is entered for the first time, with a NEXT$^*$ request, it
will always produce a solution. We denote such a 
proper
visit a 
\emph{\first\ visit}.

\begin{table}
  \centering
  \begin{tabular}{|ll|l|}
    \hline
    node type&
               type of visit&visits of children\\
    \hline
    union node & \first  & \first(1)\\
             & proper & proper(1) \\&& or
                        dummy(1)+\first(2)
                        \\&& or
                        proper(2)
    \\
             &dummy & dummy(2)\\
    \hline
    product node & \first  & \first(1)+\first(2)\\
             & proper & proper(2) \\&& or
                        dummy(2)+proper(1)+\first(2)
    \\
             &dummy & dummy(1)+dummy(2)\\
    \hline
  \end{tabular}
  \caption{The visits of the children (child 1 or child 2) that are spawned by a
    visit of a node, according to the type of visit. In this table, ``proper''
    denotes a proper visit that is not a \first\ 
    visit.}
  \label{tab:spawn}
\end{table}

Table~\ref{tab:spawn} shows the visits to the child nodes that are
caused by each type of visit. This information can be directly
extracted from the flow graphs of
Figure~\ref{fig:flow-python}.

\begin{lemma}\label{proper-visit}
    Let $K$ be a node that is associated to a subtree $A$.
  We consider a visit of $K$, from the time when $K$ receives a message
  from its parent to the first time when it returns a message to its parent.
  \begin{enumerate}
  \item In a \first\  visit and in a dummy visit, the set of visited
    nodes
    forms a \cwstree\ with root~$K$.
    In total, the number $p$ of visited product nodes is $|A|-1$.
  \item In a proper visit, 
the total number $p$ of visited product nodes is at most $2(|A|-1)$.
\item Any visit is finished after visiting $O(|A|)$ nodes in total.
  \end{enumerate}
\end{lemma}

\begin{proof}
  (1) It can be directly seen in
  Table~\ref{tab:spawn} that
  dummy visits lead only to dummy visits, \first\  visits lead
  only to \first\  visits, and they follow the pattern of a \wstree.
  
  (2) We prove this by induction, 
  following the partial order defined by the expression DAG.
  As induction basis, we consider the basis nodes. They have
  $|A|=1$ and $p=0$, and the statement is clearly true.
  
  Let us now consider a union node $K$. If
  only one of its children is visited, induction works.
  The bad case is ``dummy(1)+\first(2)''. But in that case, we apply
  part (1) and get exactly $p=(|A|-1)+(|A|-1)=2(|A|-1)$ visited product
  nodes.

  When $K$ is a product node, let us denote the vertex sets associated
  to the children by $A_1$ and $A_2$, with $|A_1|+|A_2|=|A|$.
  The case ``proper(2)'' is easy: $p=1+2(|A_2|-1)\le 2(|A|-1)$.
  In the other case, ``dummy(2)+proper(1)+\first(2)'', we
  apply the inductive hypothesis for the first child and statement~(1) of the lemma twice for the
  second child, and we get the upper bound
  \begin{displaymath} p\le
    1+ 2(|A_1|-1)+ 2(|A_2|-1)
<    2(|A|-1).
  \end{displaymath}

  (3) Consider the tree of recursive node visits, with repetitions allowed:
  Every node appears as often as it is visited.   Removing the
  product nodes  decomposes the tree into components. Each
  component consists purely of union nodes, possibly extended with
  basis nodes at the leaves. If there are $p$ visits to product nodes,
  the number of resulting components is at most $4p+1$, since every
  product node has at most three arcs to its child visits and one arc
  to its parent.

  We now use the property of the expression DAG that it contains at
  most 8 successive levels of union nodes without intervening product
  nodes. Thus, even if we generously allow
   every
   union node to cause 3 visits of its children,
the number of visited union nodes in a component is bounded by a constant.
Since the number of components is $O(p)$,
  the total number of visits is bounded by $O(p)$.
By (1) and (2), $p=O(|A|)$, and the claim follows.  
\end{proof}

\begin{theorem}
  The \textsc{Python} program ENUM1
  of Section~\ref{enumeration}
  enumerates
   the \mds s of a
   tree with linear delay, after linear setup time.
   After the last solution, the algorithm terminates in linear time.
\end{theorem}
\begin{proof}
  This follows from Lemma~\ref{proper-visit}:
  Every solution is produced by a proper visit of the target node.
  After the last solution, there is a single dummy visit.
\end{proof}


A third algorithm ENUM3, similar in spirit to
the \textsc{Python} program but without dummy visits,
is given in
Appendix~\ref{variation-enum}\ifARXIV\else of the
preprint~\cite{arxiv}\fi.




\fi
\section{Upper Bounds}
\label{upper}

We will now
 use the counting algorithm of Section~\ref{counting}
to analyze the possible numbers of
minimal dominating sets among the trees with $n$ vertices:

The following iteration 
computes the set $\mathcal{V}_n$ of all possible vectors 
of rooted trees of $n$ vertices.
\begin{align}
\label{M1}
\mathcal{V}_1&:=\{(0,1,0,0,0,1)\}
\\
\mathcal{V}_n&:=\bigcup_{1\le i<n} \mathcal{V}_i \circ
\mathcal{V}_{n-i},\text{ for $n\ge 2$}
\label{induction}
\end{align}
The operation $\circ$ in \eqref{induction} is the elementwise composition
using $\star$ applied to sets of vectors:
$$V \circ V' = \{\,x\star y\mid x\in V, y\in V'\,\}
$$
 The largest number $M_n$ of
minimal dominating sets among the trees with $n$ vertices is then
directly obtained by the formula
\ifsoda
\else
\begin{align}
\label{Mn}
  M_n&=  \max\{\,\bar M(v) \mid v\in \mathcal{V}_n\}
=  \max\{\,G+S+d+p \mid(G,S,L,d,p,f)\in \mathcal{V}_n\}
\end{align}
\fi
%
%
Table~\ref{result} below tabulates the results of this computation, and
Figure~\ref{fig:tab} represents it graphically.
We will discuss the results in Section~\ref{upper1}.

\iffull
Incidentally, with the same recursion, we also determined the \emph{smallest}
number of \mds s that a tree can have: it is 2, for trees with at
least 2 vertices, as witnessed by the star $K_{1,n-1}$.
It is easy to see that there must always be at least 2 \mds s: A tree
is a bipartite graph, and in a connected bipartite
without isolated vertices, each color class
forms a \mds.
\fi

\subsection{Majorization}
The last column in Table~\ref{result} reports the sizes of
the sets $\mathcal{V}_n$.
 These sets get very large, and it is
 advantageous 
 to remove vectors 
 that cannot contribute to trees with the maximum number of \mds s.

If the elementwise order
$$(G_1,S_1,L_1,d_1,p_1,f_1)\ge
(G_2,S_2,L_2,d_2,p_2,f_2)$$
holds for two vectors in $\mathcal{V}_i$, we can obviously omit
$(G_2,S_2,L_2,d_2,p_2,f_2)$
from $\mathcal{V}_i$ without losing the chance to find the
largest number of minimal dominating sets.
This is true because the operation~$\star$ is monotone in both arguments.
We say that
$(G_1,S_1,L_1,d_1,p_1,f_1)$ \emph{majorizes}
$(G_2,S_2,L_2,d_2,p_2,f_2)$.
(Normally, we would call this relation \emph{dominance}, but since
we are using {dominating sets} already with a
graph-theoretic meaning, we have chosen this alternative term.)

A more widely applicable majorization rule is obtained by observing that there is a
partial order of \emph{preference} between
the categories:
\begin{equation}
  \label{eq:po}
\textbf{G}>\textbf{S}>\textbf{L}
\text{ and }
\textbf{d}>\textbf{p}
\end{equation}
This means, for example, that
\textbf{G} is less restrictive than \textbf{S} in the following sense:
Consider a \mds\ for $T$, whose intersection with a subtree $A$ is
of category \textbf{S}.
Replacing this partial solution inside $A$ by any other partial
solution of category \textbf{G} will lead to a valid \mds.
As a consequence,
replacing
a partial solution $D$ of category \textbf{S} by
a partial solution of category \textbf{G}
in the subtree $A$ cannot reduce the
number of \mds s that can be built
by extending $D$ to the whole tree~$T$.

A formal proof of this claim
is based on the fact that the $\star$-operation is monotone in both
arguments
with respect to the partial order \eqref{eq:po}.
It can be checked in Table~\ref{tab:combine}
that, for example, $\textbf{G}\star B$
is at least as good as $\textbf{S}\star B$ according to the partial order,
or that $A\star\textbf{d}$
is always at least as good as $A\star\textbf{p}$. In this comparison, any
result category is of course preferable to the case~``$-$'' when no
valid solution
is built.
Also, changing a category to a more preferred category will never change a
final category (which is counted as a solution) to a non-final one.

As a consequence, if, for instance, we subtract 1 from $S$ and add 1 to $G$,
the new vector
$(G+1,S-1,L,d,p,f)$ ought to majorize the original vector
$(G,S,L,d,p,f)$, even though
the elementwise comparison fails.
An easy way to accommodate these more powerful majorization rules is to
transform
 the vectors
$(G,S,L,d,p,f)$ into
$$(G,\ G+S,\ G+S+L,\ d,\ d+p,\ f)$$ before comparing them elementwise.
We denote this wider majorization criterion by the
symbol $\succeq$, and define
\begin{multline*}
(G_1,S_1,L_1,d_1,p_1,f_1)
\succeq
(G_2,S_2,L_2,d_2,p_2,f_2)
\iff\\
(G_1, G_1+S_1, G_1+S_1+L_1, d_1, d_1+p_1, f_1) \ge
(G_2, G_2+S_2, G_2+S_2+L_2, d_2, d_2+p_2, f_2),
\end{multline*}
where the comparison on the right-hand-side is just the elementwise
comparison between 6-tuples.

We summarize our considerations in the following lemma
\begin{lemma}\label{majorize}
  \begin{enumerate}
  \item 
  If 
 $v\succeq v'$
and
 $w\succeq w'$
then $v \star  w \succeq v'\star w'$.
  \item 
  If
 $v\succeq v'$, then $\bar M(v)\ge\bar M(v')$.
  \item 
 If
 $v\succeq v'$ holds for two vectors $v,v'\in \mathcal{V}_i$, we may
 remove
$v'$ from $\mathcal{V}_i$ without changing the sizes $M_n$ of
the largest \mds s found
 in the recursion \thetag{\ref{M1}--\ref{induction}}
and~\eqref{Mn}.
  \end{enumerate}
\end{lemma}
\begin{proof}
The first two items are a straightforward calculation.

To see the third claim,  
we introduce the \emph{majorized hull} of a set
$P\subseteq \mathbb{R}^6_{\ge0}$,
 denoted by $\hull( P)$: It is the set of all nonnegative 6-vectors that are
majorized by some vector in $P$ according to the relation~$\succeq$:
$$
\hull (P) := \{ x \in  \mathbb{R}^6_{\ge0}\mid
x\preceq y\text{ for some $y\in P$}\,\}
$$

Algebraically, the justification for the reduction to the majorized hull
comes from the following equations.
\begin{align}
  \label{eq:hull}
  \hull(P\circ Q)& =\hull(   \hull(P)\circ  \hull( Q))
\\
  \label{eq:hull-union}
  \hull(P\cup Q)& =\hull(   \hull(P)\cup  \hull( Q))
\end{align}
Equation \eqref{eq:hull} comes directly from part~1
of the
lemma, 
and \eqref{eq:hull-union} follows from the transitivity of~$\preceq$.

Reading the equations \thetag{\ref{eq:hull}--\ref{eq:hull-union}}
from left to right, they say: If we are
interested only in the hull of a ``product'' $P\circ Q$ or a union
$P\cup Q$, we might as well take the hull of the parts $P$ and $Q$
before performing the operation.
Since the set $\mathcal{V}_n$ in the iteration \eqref{induction} is
built up from smaller sets $\mathcal{V}_i$ by $\circ$ and $\cup$
operations,
this justifies the application of the hull operation at every level,
proving part~3 of the lemma.
\end{proof}

\subsection{The convex hull}
\label{sec:hull}

We can further reduce the size of the point sets by taking the convex
hull,
$\conv(P)$.
We combine the convex hull and the majorized hull in one operation
$
\hullp(P)
=
\conv(\hull(P))
=
\hull(\conv(P))
$, which we call the \emph{majorized convex hull}.
\iffull
The majorized convex hull can also be formed by taking the convex
hull together with the
rays in directions
$(-1,1,0,0,0,0)$,
$(0,-1,1,0,0,0)$,
$(0,0,0,-1,1,0)$,
as well as the coordinate directions
$(0,0,-1,0,0,0)$,
$(0,0,0,\allowbreak 0,-1,0)$, and
$(0,0,0,0,0,-1)$,
and clipping the result to the nonnegative orthant.

\fi
We have the same properties as for the majorized hull:
\begin{lemma}
\begin{align}
  \label{eq:conv-circ}
  \conv(P\circ Q) &=\conv(   \conv(P)\circ  \conv( Q))
\\
  \label{eq:conv-union}
  \conv(P\cup Q) &=\conv(   \conv(P)\cup  \conv( Q))
\\
  \label{eq:hullp-circ}
  \hullp(P\circ Q) &=\hullp(   \hullp(P)\circ  \hullp( Q))
\\
  \label{eq:hullp-union}
  \hullp(P\cup Q) &=\hullp(   \hullp(P)\cup  \hullp( Q))
\end{align}
\end{lemma}
\begin{proof}
To prove~\eqref{eq:conv-circ}, we
first prove
\begin{equation}\label{convex-inclusion}
  \conv(P\circ Q) \supseteq   \conv(P)\circ  \conv( Q),
\end{equation}
using the fact
 that the function
$\star\colon
 \mathbb{R}^6_{\ge0} \times
 \mathbb{R}^6_{\ge0} \to
 \mathbb{R}^6_{\ge0}$
is bilinear.
An element formed from two convex combinations on the right-hand side is of the form
\begin{align*}
\sum_i{\mu_ip_i} \star   
\sum_j{\nu_jq_j} =
\sum_i\sum_j\mu_i\nu_j(p_i\star q_j),
\end{align*}
with $\sum_i\sum_j\mu_i\nu_j=1$, and is hence an element of
$\conv(P\circ Q)$.
From \eqref{convex-inclusion}, the inclusion
$  \conv(P\circ Q) \supseteq \conv(   \conv(P)\circ  \conv( Q))$
follows by a standard convexity argument, and the reverse conclusion
is an easy consequence of the inclusion $P\subseteq \conv(P)$.

 Equation \eqref{eq:conv-union} is standard, and
\eqref{eq:hullp-circ} and
\eqref{eq:hullp-union} follow from combining the equations
\thetag{\ref{eq:conv-circ}--\ref{eq:conv-union}} for the convex hull with
the equations \thetag{\ref{eq:hull}--\ref{eq:hull-union}} for the
majorized hull.
\end{proof}

We are interested in the maximum total $\bar M$, which is a linear
function, and hence the convex hull is sufficient.
Equation~\eqref{eq:conv-circ} tells us that to compute
$\conv(P\circ Q)$, it is sufficient to compute $v\star w$ for the
vertices of $P$ and $Q$ and take the convex hull.

\begin{table*}[pht]
  \begingroup
  \def\extraskip{0pt}
\ifsoda
\else
\ifTALG \small \def\extraskip{-1.16pt}
\fi
\newcount\mooo
\mooo = 0

\def \haa #1 #2 #3 #4 {%
\ifnum \mooo=0#1 #4 
\fi
\ifnum 0=0#1
\let \next=\relax \fi
\next}

\def \maa{\global\advance\mooo by 1
\let\next=\haa
\haa
1 x x 1
2 1.41421356237  2 1
3 1.25992104989  2 2
4 1.41421356237  4 2
5 1.31950791077  4 4
6 1.41421356237  8 5
7 1.36873810664  9 9
8 1.41421356237  16 13
9 1.38702322584  19 19
10 1.41421356237  32 32
11 1.40157620021  41 39
12 1.41421356237  64 73
13 1.40739771128  85 85
14 1.41421356237  128 144
15 1.4120981512   177 176
16 1.41421356237  256 279
17 1.41397457412  361 337
18 1.41421356237  512 492
19 1.41553085871  737 612
20 1.41421356237  1024 841
21 1.41608793849  1489 1055
22 1.41421356237  2048 1320
23 1.41656252138  3009 1641
24 1.41421356237  4096 1969
25 1.41666558385  6049 2435
26 1.41421356237  8192 2805
27 1.41675632056  12161 3456
28 1.41421356237  16384 3871
29 1.41670718071  24385 4656
30 1.41421356237  32768 5329
31 1.41666501244  48897 6227
32 1.41449859436  65960 7248
33 1.41657202788  97921 8436
34 1.41526678247  134432 9719
35 1.41648981353  196097 11277
36 1.41569656429  272224 12878
37 1.41639156077  392449 14890
38 1.41609068088  551392 16931
39 1.41630342193  785409 19088
40 1.41634892846  1113808 22214
41 1.4162126408   1571329 24075
42 1.41658315524  2249920 28344
43 1.41613031645  3143681 30029
44 1.41668758344  4529600 35068
45 1.41605019185  6288385 36809
46 1.41678485046  9119680 42438
47 1.41597689194  12578817 44773
48 1.416828082  18332576 50902
49 1.41590722737  25159681 54417
50 1.41686791093  36852608 61859
51 1.41584303009  50323457 66246
0 . . .
}

\def \hbb n = #1 #2 #3 #4 {%
\ifnum \mooo=0#1 #4
\fi
\ifnum 0=0#1
\let \next=\relax \fi
\next}

\def \mbb{\let\next=\hbb
\hbb 
n = 1 x x 1
n = 2 1.41421356237 2 1
n = 3 1.25992104989 2 2
n = 4 1.41421356237 4 4
n = 5 1.31950791077 4 7
n = 6 1.41421356237 8 13
n = 7 1.36873810664 9 24
n = 8 1.41421356237 16 45
n = 9 1.38702322584 19 85
n = 10 1.41421356237 32 159
n = 11 1.40157620021 41 308
n = 12 1.41421356237 64 588
n = 13 1.40739771128 85 1180
n = 14 1.41421356237 128 2326
n = 15 1.4120981512 177 4753
n = 16 1.41421356237 256 9591
n = 17 1.41397457412 361 19793
n = 18 1.41421356237 512 40638
n = 19 1.41553085871 737 84641
n = 20 1.41421356237 1024 176255
n = 21 1.41608793849 1489 369635
n = 22 1.41421356237 2048 775935
n = 23 1.41656252138 3009 1634901
n = 24 1.41421356237 4096 3451490
n = 25 1.41666558385 6049 7303232
n = 26 1.41421356237 8192 15481738
n = 27 1.41675632056 12161 32868146
n = 0 . . .
}

  \centering
  \begin{tabular}{r|lrrrr}%
$n$& $\sqrt[n]{M_n}$&$M_n$&
 $\#\hullp\!(\mathcal{V}_n)$&
 $\#\hull(\mathcal{V}_n)$&
 $|\mathcal{V}_n|$\\
\hline    
1&1&1&1&\maa&\mbb\\[\extraskip]
2&1.41421356237310&2&1&\maa&\mbb\\[\extraskip]
3&1.25992104989487&2&2&\maa&\mbb\\[\extraskip]
4&1.41421356237310&4&2&\maa&\mbb\\[\extraskip]
5&1.31950791077289&4&4&\maa&\mbb\\[\extraskip]
6&1.41421356237309&8&3&\maa&\mbb\\[\extraskip]
7&1.36873810664220&9&6&\maa&\mbb\\[\extraskip]
8&1.41421356237310&16&7&\maa&\mbb\\[\extraskip]
9&1.38702322584422&19&11&\maa&\mbb\\[\extraskip]
10&1.41421356237310&32&14&\maa&\mbb\\[\extraskip]
\hline
11&1.40157620020641&41&17&\maa&\mbb\\[\extraskip]
12&1.41421356237309&64&24&\maa&\mbb\\[\extraskip]
13&1.40739771128108&85&26&\maa&\mbb\\[\extraskip]
14&1.41421356237309&128&30&\maa&\mbb\\[\extraskip]
15&1.41209815120249&177&30&\maa&\mbb\\[\extraskip]
16&1.41421356237310&256&36&\maa&\mbb\\[\extraskip]
17&1.41397457411881&361&39&\maa&\mbb\\[\extraskip]
18&1.41421356237309&512&51&\maa&\mbb\\[\extraskip]
19&1.41553085871039&737&47&\maa&\mbb\\[\extraskip]
20&1.41421356237310&1024&66&\maa&\mbb\\[\extraskip]
\hline
21&1.41608793848702&1489&58&\maa&\mbb\\[\extraskip]
22&1.41421356237310&2048&74&\maa&\mbb\\[\extraskip]
23&1.41656252137841&3009&62&\maa&\mbb\\[\extraskip]
24&1.41421356237309&4096&93&\maa&\mbb\\[\extraskip]
25&1.41666558384650&6049&75&\maa&\mbb\\[\extraskip]
26&1.41421356237310&8192&111&\maa&\mbb\\[\extraskip]
27&1.41675632056381&12161&87&\maa&\mbb\\[\extraskip]
28&1.41421356237309&16384&119&\maa&\mbb\\[\extraskip]
29&1.41670718070637&24385&102&\maa&\mbb\\[\extraskip]
30&1.41421356237310&32768&125&\maa&\mbb\\[\extraskip]
\hline
31&1.41666501243844&48897&116&\maa&\mbb\\[\extraskip]
32&1.41449859435768&65960&123&\maa&\mbb\\[\extraskip]
33&1.41657202787702&97921&129&\maa&\mbb\\[\extraskip]
34&1.41526678247498&134432&130&\maa&\mbb\\[\extraskip]
35&1.41648981352598&196097&146&\maa&\mbb\\[\extraskip]
36&1.41569656428574&272224&151&\maa&\mbb\\[\extraskip]
37&1.41639156076937&392449&177&\maa&\mbb\\[\extraskip]
38&1.41609068088382&551392&166&\maa&\mbb\\[\extraskip]
39&1.41630342192653&785409&193&\maa&\mbb\\[\extraskip]
40&1.41634892845829&1113808&184&\maa&\mbb\\[\extraskip]
\hline
41&1.41621264079532&1571329&209&\maa&\mbb\\[\extraskip]
42&1.41658315523612&2249920&217&\maa&\mbb\\[\extraskip]
43&1.41613031644569&3143681&212&\maa&\mbb\\[\extraskip]
44&1.41668758343879&4529600&238&\maa&\mbb\\[\extraskip]
45&1.41605019185075&6288385&220&\maa&\mbb\\[\extraskip]
46&1.41678485046458&9119680&240&\maa&\mbb\\[\extraskip]
47&1.41597689193916&12578817&233&\maa&\mbb\\[\extraskip]
48&1.41682808199910&18332576&273&\maa&\mbb\\[\extraskip]
49&1.41590722737106&25159681&260&\maa&\mbb\\[\extraskip]
50&1.41686791092506&36852608&287&\maa&\mbb\\[\extraskip]
\hline
51&1.41584303009330&50323457&264&\maa&\mbb\\[\extraskip]
52&1.41685798299446&73955200&293&\maa&\mbb\\[\extraskip]
\hline
  \end{tabular}

\fi
\endgroup
\ifTALG \smallskip\fi
  \caption{The maximum number $M_n$ of \mds s of a tree with $n$
    vertices.
  $\#\hull(\mathcal{V}_n)$ denotes the number of generating vertices
of $\hull(\mathcal{V}_n)$ (the non-majorized
  vertices of $\mathcal{V}_n$), and
  $\#\hullp(\mathcal{V}_n)$ is the number of extreme  non-majorized
  vertices in
 $\hullp(\mathcal{V}_n)$.}
  \label{result}
\end{table*}

\begin{figure*}[b]
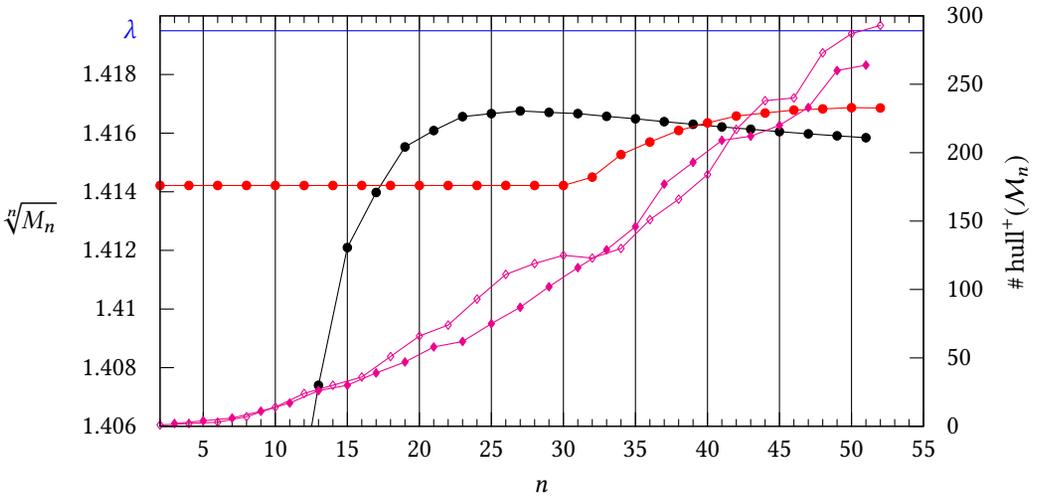

  \centering
\hrule height 0pt
\vskip -2cm
\hrule height 0pt
\input res-figure
  \Description{A plot of the $M_n$ values from Table~\ref{result}}
  \caption{The $n$-th root of the maximum number $M_n$ of \mds s of trees with $n$
    vertices.
Even and odd values of $n$ (red and black dots) behave differently.
\iffull
The pink curves through the diamonds
show the growth of the convex non-majorized
hulls, 
$\hullp(\mathcal{V}_n)$.
Again, even and odd values of~$n$ behave differently.
\fi}
  \label{fig:tab}
\end{figure*}

\subsection{The upper bound for trees of a given size}
\label{upper1}

We have carried out the iteration \eqref{induction} for calculating
$M_n$, both with the {majorized hull}, $\hull(\mathcal{V}_n)$,
and the {majorized convex hull}, $\hullp(\mathcal{V}_n)$.
The results are presented in 
Table~\ref{result} and Figure~\ref{fig:tab}.
 Figure~\ref{fig:tab} shows clearly that the trees with even and odd $n$ behave
differently.
For a while,
$\sqrt[n]{M_n}$ for
 the even trees remains constant at $\sqrt 2$,
which comes from
repeating the tree with two vertices, while the odd trees
rise from a low
 start. They overtake the even trees for $n=19$ and reach a local
maximum at $n=27$. The corresponding value 
$\sqrt[27]{12161}\approx 1.416756$ was the best lower bound on
$\lambda$ known so far, due to Krzywkowski~\shortcite{k-tmmds-13}.
The optimal tree with 27 vertices, which has 12161 \mds s, consists of
two snowflakes and an additional vertex that is attached to the
centers of the two snowflakes.
We suspect that Krzywkowski must have run a program like ours to come
up with this tree.
In Figure~\ref{fig:tab} it is also apparent that the values
stay well below the true bound~$\lambda$. There is no way how one
could have guessed the limiting behavior from these numbers, even if
the range of sizes $n$ could be extended.
\iffull
In fact, \emph{all} optimal trees of odd order that are reported in
the table have the same ``double-snowflake'' structure.
The number of arms of the snowflakes must be varied to reach the
desired number of vertices; the arms are distributed as
equally as possible to the two snowflakes. (For $n\le 7$, these trees
degenerate to paths.)
At $n=32$, the even values start to increase, leading to new records
 for $n\ge46$, while the odd values continue to decrease.
All optimal trees of even order $n$ that we found for $n\ge32$ have a
similar structure, see Figure~\ref{optimal44}. They consist of two double-snowflakes of odd order
$n_1$ and $n_2$ with $n_1+n_2=n$ and
$n_1$ and $n_2$ as close together as possible, connected by an edge
between two snowflake centers. When there is a choice, the center
of the smaller snowflake is used as an endpoint of the connecting edge.
The trees of this pattern reach their local maximum at
$\sqrt[50]{M_{50}}
=\sqrt[50]{ 36\,852\,608} 
\approx 1.41686791$. Beyond this size, they decline, and at some
point, trees with three, five, or six snowflakes will probably begin to
take the lead.

\begin{figure}[htb]
  \centering
  \includegraphics{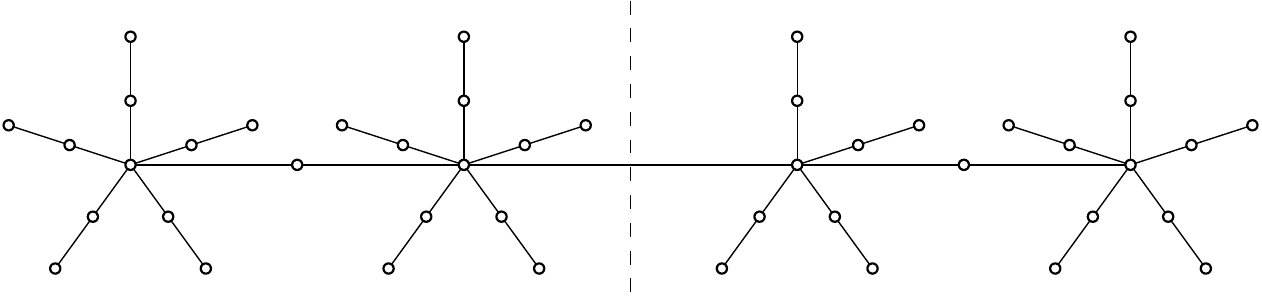}
  \Description{The structure of an optimal tree, as described in the text}
  \caption{An optimal tree with 44 vertices. The left and the right
    half is an optimal tree with 23 and 21 vertices, respectively.}
  \label{optimal44}
\end{figure}

The even optimal trees with $2^{n/2}$ \mds s are far from
unique:
One can start with an arbitrary tree on $n/2$ vertices and add a new
leaf adjacent to each vertex.
We did not check whether the other classes of optimal trees that we
found are unique.

\fi

We can now describe how Part 2 of Theorem~\ref{sizes} is obtained. 
For $n \ge 38$,
we construct a tree with
at least $0.649748 \cdot \lambda^n$ \mds s with the help of
 the supermultiplicativity property of
Observation~%
\ref{obse}(\ref{super-function}) as follows.
If $n\ge37$ and $n$ is congruent to $1,2,\ldots,13$ modulo 13,
we combine the optimum tree of size
$0, 14, 2, 16, 4, 18, 6, 20, 8, 35, 10, 37, 12$
from Table~\ref{result}
with a record tree
$\mathrm{RT}_{13k+1}$
 from the end of Section~\ref{lower} of appropriate size. 
(The factor 0.649748 in the claim is restricted by the
tree of size 37 in this list.)  
For $n< 37$, the trees
in Table~\ref{result} do the job.
%


\iffull
\paragraph{Implementation details and program runs.}
\fi

\def\implementationdetails{%
The version of the program which uses only
 the majorized hull for pruning points is very simple and
 did not pose any challenges.
We used a pairwise comparison of all generated elements to 
remove majorized vectors. 
The program was written in the \textsc{Python} programming language
and has less than 100 lines, including rudimentary code to print optimal trees.
As the fifth column of Table~\ref{result} shows, the
number of non-majorized vectors grows fast.

Therefore, we used the convex hull to further reduce the number of points
that need to be stored and processed.
For the convex-hull computations, 
we tested for each generated vector whether it is
a convex combination of the remaining vectors, and deleted it in case
of a positive answer.
This test can be formulated as a linear programming problem.
We wrote our program for the
 mathematics software system \textsc{sage}\footnote
{\url{http://www.sagemath.org/}}, which provides straightforward access
to linear programming.
 We used the default solver
GLPK that is installed with \textsc{sage}.
As the fourth column shows, using the convex hull leads to a
substantial reduction of the number of vertices that need to be stored
and processed, allowing us to carry the computation further than
without the convex-hull computations,
and we managed to compute the values up to $M_{52}$.
The number of non-majorized convex hull vertices appears to increase quadratically with
$n$.
This means that the number of points that are generated in \eqref{induction} and
subjected to the redundancy test in the computation of each new entry
$M_n$ grows like $n^5$. 
The calculations
ran for several months.


We must concede that, due to the error-prone nature of
floating-point computations, the results for $n=52$ cannot be considered as
totally reliable. It is conceivable that an extreme vertex is
erroneously pruned because of numerical errors in the solution of the
linear programs, leading to missing trees.
  However, as the dimension
of the problem and the involved numbers are not very big, this is
probably not an issue.  
(In contrast, for the results
\iffull
 that we
will mention below
\else
mentioned
\fi
in Section~\ref{upper2}, we undertook the effort to certify
the linear-programming results a posteriori.)

\iffull
The given values of $M_n$ are certainly valid as a lower bounds, as
each
value comes from a computation that represents an actual tree.
For $n\le 51$, where a
number is reported in the fifth column,
the results are not subject to these reservations, because they are confirmed
by the reliable calculation without convex-hull computation.

 \fi
}

\iffull\implementationdetails\fi


\subsection{Characterization of the growth rate}
\label{upper2}

Since the sequence $M_n$ is supermultiplicative (Observation~%
\ref{obse}(\ref{super-function})) and bounded by an exponential
function $M_n\le 2^n$, it follows from Fekete's lemma
that the limit
\begin{equation}
  \label{eq:lambda}
  \lambda^* := \lim_{n\to \infty} \sqrt[n]{M_n}
\end{equation}
exists and that 
\begin{equation}
  \label{eq:strict-bound}
 M_n \le (\lambda^*)^n. 
\end{equation}
In contrast to the previous parts, we now denote the growth rate
by $\lambda^*$,
and we will use  $\lambda$ for a generic ``test value'', not necessarily
the
correct growth rate.
The following statement provides a characterization of%
~$\lambda^*$.

\begin{proposition}\label{characterize}
The growth constant $\lambda^*$ equals the smallest
 the value
 $\lambda$ for which there exists a bounded
convex 
 set $P$ with $P=\hullp P$ such
that
\begin{equation}\label{basis}
 (0,1,0,0,0,1)/\lambda\in P
\end{equation}
and 
\begin{equation}\label{closed}
  P\circ P \subseteq P.
\end{equation}
\end{proposition}
\begin{proof}
First we show that the statement does not change if we omit the
condition that
$P$ is convex 
 and that 
$P=\hullp(P)$:
If this condition is not fulfilled by some set $P$,
we can simply replace
$P$ with $\hullp(P)$.
This will of course not affect~\eqref{basis}, and
by \eqref{eq:hullp-circ},
 taking the majorized convex hull of $P$ does not
 invalidate the condition 
 $P\circ P \subseteq P$. 

We can write down the smallest set $P$ fulfilling the required
  properties~\eqref{basis} and~\eqref{closed}. It is 
\begin{equation}
  \label{eq:P0}
P_0 := \bigcup_{n\ge 1} \mathcal{V}_n/\lambda^n.  
\end{equation}
Let us see why this is true.
By assumption~\eqref{basis},  $\mathcal{V}_1/\lambda =
\{
 (0,1,0,0,0,1)/\lambda\}$ must be contained in $P_0$.
Let us now consider
a vector
$v\in \mathcal{V}_n$. It must be the result
 $w\star w'$ for some vectors
$w\in \mathcal{V}_i$ and
$w'\in \mathcal{V}_j$ with $i+j=n$.
If we assume by induction that
$w/\lambda^i$ and $w'/\lambda^j$ are in $P_0$, we conclude from 
\eqref{closed} that
$w/\lambda^i\star w'/\lambda^j=v/\lambda^n$ is also in $P_0$.

We will now prove the proposition through a sequence of equivalent statements:
\begin{align}
  \text{\ifsoda\else \let\next\relax\fi
  bo\next{unded $P$ exists for $\lambda$}}
 &\iff \label{1}
\text{$P_0$ is bounded}\\
 &\iff  \label{2}
\text{the sequence $\|\mathcal{V}_n\|_1/\lambda^n$ is bounded}\\
 &\iff  \label{3}
\text{the sequence $M_n/\lambda^n$ is bounded}\\
 &\iff  \label{4}
\lim_{n\to\infty} \sqrt[n]{M_n/\lambda^n} \le 1\\
 &\iff \lambda^*/\lambda \le 1 \iff \lambda \ge \lambda^*
 \label{5}
\end{align}
The equivalence between 
the first and the last statement
is the claim of the proposition.

The equivalence
\eqref{1} has already been shown above.
In  \eqref{2}, we have decided to use the $l_1$ norm for expressing boundedness:
 $\|\mathcal{V}_n\|_1 := \max \{\,
\|v\|_1 
\mid v \in\mathcal{V}_n\,\}$.
The equivalence follows from the definition \eqref{eq:P0} of $P_0$.
When proceeding to \eqref{3}, we are replacing the $l_1$-norm
$\|v\|_1 $
 by the
function $\bar M(v)$, which sums only 4 of the 6 entries of $v$.
To
justify this change, we
show that it does not change the notion of boundedness. 
it is sufficient to prove the following relation:
\begin{equation}\label{sandwich}
 M_n \le  \|\mathcal{V}_n\|_1 \le M_{n+3}
\end{equation}
The left inequality is trivial, because
$G+S + d+f \le G+S+L + d+p+f$.
The converse inequality is not true, because the categories \textbf{L}
and \textbf{p} are not counted for $\bar M$. However, by appending a path
of length 3 to the root, we ensure that every partial solution,
no matter of which category, can be completed to a valid \mds\ in the
larger tree. Algebraically, this can be checked by the following
calculation:
\ifsoda
\else
\begin{align*}
v_0\star (v_0\star (v_0\star (G,S,L,d,p,f)))\hphantom)
&=
 (G + S+L, d + f, d + p, G + S + d + p, f, G + d + f)\\
\bar M(v_0\star (v_0\star (v_0\star (G,S,L,d,p,f))))
& =
2G + 2S + L + 2d + p + 2f \ge \|(G,S,L,d,p,f)\|_1
\end{align*}
\fi
This means that, for every tree with $n$ nodes and vector~$v$, there
is a tree with $n+3$ nodes and vector $v'$ such that $\bar M(v')\ge
\|v\|_1$.
This establishes the right inequality of~\eqref{sandwich}.

The equivalence between \eqref{3} and
\eqref{4} is obvious except in the borderline case when the limit
$\lim_{n\to\infty} \sqrt[n]{M_n/\lambda^n}$ equals $1$, so let us postpone
this case for the moment.
The remaining steps till \eqref{5} are straightforward in view of 
the known value of the limit~\eqref{eq:lambda}.

Let us return to the borderline case $\lambda=\lambda^*$.
In this case, \eqref{eq:strict-bound} tells us that
 $M_n/\lambda^n\le 1$ for all $n$, and thus the equivalence
between
\eqref{3} and \thetag{\ref{4}--\ref{5}} holds also in this case.
\end{proof}

\subsection{Automatic determination of the growth factor}
\label{automatic}

The property of
$P$ that is required in
Proposition~\ref{characterize}
 is monotone in the sense that if it can be fulfilled for some
 $\lambda$, the same set $P$ will also work
 for all larger values of $\lambda$.
This holds because since $P$ contains its majorized hull,
and therefore property~\eqref{basis} remains fulfilled.
This opens the way for a semi-automatic experimental way
to search for the correct growth factor~$\lambda^*$.
\goodbreak
\begin{enumerate}
\item Choose a trial value $\lambda$, and set $Q := 
\{(0,1,0,0,0,1)/\lambda\}$.
\item Form the set $Q^{2} := Q \circ Q$ of all pairwise products of $Q$.
\item Compute $P := \hullp(Q\cup Q^{2})$.
\item Let $Q$ be the set of non-majorized vertices of $P$.
\item Repeat from
Step~2 until the process converges or diverges.
\label{diverge}
\item If divergence occurs, $\lambda$ was chosen too small, and a
  larger value must be tried. In case of convergence, try a smaller value.
\end{enumerate}

In practice, divergence in Step~\ref{diverge} manifests itself in an exponential
growth of the vector entries and is easy to detect once it sets in.
The trees corresponding to the vectors which are ``responsible'' for
the divergence have more than $\lambda^n$ \mds s. By looking at such trees, we got the
idea for the lower-bound construction in Section~\ref{lower}.
As it turned
out,
we were lucky, and this construction gave the correct value of $\lambda^*$.

\begin{table}[ptb]
  \small
  \centering
\noindent
\begin{minipage}[t]{0.47\linewidth}
\begin{tabular}[t]{@{}l@{}l}
$v_{1}=v_{1}\star v_{32}$&${} = (0.9, 0, 0, 0, 0, 0) $\\
$v_{2}$&${} = (0, 1, 0, 0, 0, 1) \lambda^{-1}$\\
$v_{3}=v_{2}\star v_{2}$&${} = (1, 0, 0, 1, 0, 0) \lambda^{-2}$\\
$v_{4}=v_{2}\star v_{3}$&${} = (0, 1, 1, 1, 0, 1) \lambda^{-3}$\\
$v_{5}=v_{2}\star v_{4}$&${} = (1, 1, 0, 1, 1, 1) \lambda^{-4}$\\
$v_{6}=v_{4}\star v_{3}$&${} = (0, 1, 3, 3, 0, 1) \lambda^{-5}$\\
$v_{7}=v_{2}\star v_{5}$&${} = (1, 1, 1, 2, 0, 2) \lambda^{-5}$\\
$v_{8}=v_{2}\star v_{6}$&${} = (1, 3, 0, 1, 3, 3) \lambda^{-6}$\\
$v_{9}=v_{6}\star v_{3}$&${} = (0, 1, 7, 7, 0, 1) \lambda^{-7}$\\
$v_{10}=v_{7}\star v_{3}=v_{4}\star v_{5}$&${} = (2, 1, 3, 6, 0, 2) \lambda^{-7}$\\
$v_{11}=v_{2}\star v_{8}$&${} = (3, 1, 1, 4, 0, 4) \lambda^{-7}$\\
$v_{12}=v_{2}\star v_{9}$&${} = (1, 7, 0, 1, 7, 7) \lambda^{-8}$\\
$v_{13}=v_{9}\star v_{3}$&${} = (0, 1, 15, 15, 0, 1) \lambda^{-9}$\\
$v_{14}=v_{6}\star v_{5}=v_{10}\star v_{3}$&${} = (4, 1, 7, 14, 0, 2) \lambda^{-9}$\\
$v_{15}=v_{11}\star v_{3}=v_{4}\star v_{8}$&${} = (6, 1, 3, 12, 0, 4) \lambda^{-9}$\\
$v_{16}=v_{2}\star v_{12}$&${} = (7, 1, 1, 8, 0, 8) \lambda^{-9}$\\
$v_{17}=v_{2}\star v_{13}$&${} = (1, 15, 0, 1, 15, 15) \lambda^{-10}$\\
$v_{18}=v_{2}\star v_{14}$&${} = (2, 14, 4, 5, 7, 14) \lambda^{-10}$\\
$v_{19}=v_{13}\star v_{3}$&${} = (0, 1, 31, 31, 0, 1) \lambda^{-11}$\\
$v_{20}=v_{9}\star v_{5}=v_{14}\star v_{3}$&${} = (8, 1, 15, 30, 0, 2) \lambda^{-11}$\\
$v_{21}=v_{6}\star v_{8}=v_{15}\star v_{3}$&${} = (12, 1, 7, 28, 0, 4) \lambda^{-11}$\\
$v_{22}=v_{4}\star v_{12}=v_{16}\star v_{3}$&${} = (14, 1, 3, 24, 0, 8) \lambda^{-11}$\\
$v_{23}=v_{2}\star v_{17}$&${} = (15, 1, 1, 16, 0, 16) \lambda^{-11}$\\
$v_{24}=v_{2}\star v_{19}$&${} = (1, 31, 0, 1, 31, 31) \lambda^{-12}$\\
$v_{25}=v_{2}\star v_{20}$&${} = (2, 30, 8, 9, 15, 30) \lambda^{-12}$\\
  $v_{26}=v_{19}\star v_{3}$&${} = (0, 1, 63, 63, 0, 1) \lambda^{-13}$%
\end{tabular}
\end{minipage}
\hfill
\begin{minipage}[t]{0.52\linewidth}
\begin{tabular}[t]{l@{}l}
$v_{27}=v_{20}\star v_{3}=v_{13}\star v_{5}$&${} = (16, 1, 31, 62, 0, 2) \lambda^{-13}$\\
$v_{28}=v_{21}\star v_{3}=v_{9}\star v_{8}$&${} = (24, 1, 15, 60, 0, 4) \lambda^{-13}$\\
$v_{29}=v_{6}\star v_{12}=v_{22}\star v_{3}$&${} = (28, 1, 7, 56, 0, 8) \lambda^{-13}$\\
$v_{30}=v_{4}\star v_{17}=v_{23}\star v_{3}$&${} = (30, 1, 3, 48, 0, 16) \lambda^{-13}$\\
$v_{31}=v_{2}\star v_{25}$&${} = (30, 9, 2, 32, 8, 24) \lambda^{-13}$\\
$v_{32}=v_{2}\star v_{24}$&${} = (31, 1, 1, 32, 0, 32) \lambda^{-13}$\\
$v_{33}=v_{2}\star v_{26}$&${} = (1, 63, 0, 1, 63, 63) \lambda^{-14}$\\
$v_{34}=v_{2}\star v_{27}$&${} = (2, 62, 16, 17, 31, 62) \lambda^{-14}$\\
$v_{35}=v_{26}\star v_{3}$&${} = (0, 1, 127, 127, 0, 1) \lambda^{-15}$\\
$v_{36}=v_{19}\star v_{5}=v_{27}\star v_{3}$&${} = (32, 1, 63, 126, 0, 2) \lambda^{-15}$\\
$v_{37}=v_{13}\star v_{8}=v_{28}\star v_{3}$&${} = (48, 1, 31, 124, 0, 4) \lambda^{-15}$\\
$v_{38}=v_{9}\star v_{12}=v_{29}\star v_{3}$&${} = (56, 1, 15, 120, 0, 8) \lambda^{-15}$\\
$v_{39}=v_{30}\star v_{3}=v_{6}\star v_{17}$&${} = (60, 1, 7, 112, 0, 16) \lambda^{-15}$\\
$v_{40}=v_{4}\star v_{24}=v_{32}\star v_{3}$&${} = (62, 1, 3, 96, 0, 32) \lambda^{-15}$\\
$v_{41}=v_{26}\star v_{5}=v_{36}\star v_{3}$&${} = (64, 1, 127, 254, 0, 2) \lambda^{-17}$\\
$v_{42}=v_{19}\star v_{8}=v_{37}\star v_{3}$&${} = (96, 1, 63, 252, 0, 4) \lambda^{-17}$\\
$v_{43}=v_{38}\star v_{3}=v_{13}\star v_{12}$&${} = (112, 1, 31, 248, 0, 8) \lambda^{-17}$\\
$v_{44}=v_{9}\star v_{17}=v_{39}\star v_{3}$&${} = (120, 1, 15, 240, 0, 16) \lambda^{-17}$\\
$v_{45}=v_{6}\star v_{24}=v_{40}\star v_{3}$&${} = (124, 1, 7, 224, 0, 32) \lambda^{-17}$\\
$v_{46}=v_{26}\star v_{8}=v_{42}\star v_{3}$&${} = (192, 1, 127, 508, 0, 4) \lambda^{-19}$\\
$v_{47}=v_{43}\star v_{3}=v_{19}\star v_{12}$&${} = (224, 1, 63, 504, 0, 8) \lambda^{-19}$\\
$v_{48}=v_{13}\star v_{17}=v_{44}\star v_{3}$&${} = (240, 1, 31, 496, 0, 16) \lambda^{-19}$\\
$v_{49}=v_{9}\star v_{24}=v_{45}\star v_{3}$&${} = (248, 1, 15, 480, 0, 32) \lambda^{-19}$\\
$v_{50}=v_{26}\star v_{12}=v_{47}\star v_{3}$&${} = (448, 1, 127, 1016, 0, 8) \lambda^{-21}$\\
$v_{51}=v_{48}\star v_{3}=v_{19}\star v_{17}$&${} = (480, 1, 63, 1008, 0, 16) \lambda^{-21}$\\
$v_{52}=v_{49}\star v_{3}=v_{13}\star v_{24}$&${} = (496, 1, 31, 992, 0, 32) \lambda^{-21}$
\end{tabular}
\end{minipage}
\medskip\hrule
\medskip
\begin{tabular}[t]{l@{}l}
$v_{53}=v_{24}\star v_{19}$&${} = (63, 961, 0, 63, 1922, 961) \lambda^{-23}$\\
$v_{54}=v_{52}\star v_{3}=v_{19}\star v_{24}$&${} = (992, 1, 63, 2016, 0, 32) \lambda^{-23}$\\
    $v_{55}=v_{33}\star v_{26}$&
                            ${} = (127, 3969, 0, 127, 7938, 3969) \lambda^{-27}$
\end{tabular}
\medskip
  \caption{The 55 vertices generating the polytope $P$;  $\lambda = \sqrt[13]{95} \approx 1.4195$.}
  \label{tab:P}
\end{table}

With this value of $\lambda$, we eventually determined
a set $P$ which
does the job of proving the upper bound by 
Proposition~\ref{characterize}.
It is
the set $P=\hullp(\{v_1,\ldots,v_{55}\})$ with the vectors given in
Table~\ref{tab:P},
The \emph{seed vector} $v_2=(0,1,0,0,0,1)/\lambda$ is in $P$ by
construction,
and thus the first requirement on $P$ is fulfilled.
 The vectors other than $v_1$ correspond to
actual trees, and the exponent of $1/\lambda$ given in the table is
their size. By looking at the alternate expressions 
in
the left column of the table, one can see how each tree is
constructed from smaller trees. When two trees are combined, the
exponents of $\lambda$ are added.

The ``extra'' vector $v_1=(0.9,0,0,0,0,0)$ has been chosen in the following way.
The stars of snowflakes 
from 
Section~\ref{lower}
 yield points
 $95^k(1+o(1),o(1),o(1),o(1),o(1),o(1))\lambda^{-13k-2}$
 if the vertex $a$ is chosen as the tree root. These points
converge to the vector $v_\infty:=(1,0,0,0,0,0)/\lambda
\approx(0.7044,0,0,0,0,0)$,
and this vector must belong to $P$ at least as a limit point.
On the other hand, we know from by Part~1 of Theorem~\ref{sizes} that
 no finite tree corresponds to
the point $v_\infty$, and hence, this point will never be included in
$P$ by the algorithm.
By choosing a larger rescaling $v_1$ of this vector, we move away from
the infinitely many vectors converging to $v_\infty$, hoping to
swallow them (and possibly more points) into the convex hull, thus
obtaining a smaller point set.
The value $0.9$ for the vector $v_1$ was chosen by experiment as being
close to the largest value that led to convergence.

\subsection{The necessity of irrational coordinates}
\label{sec:irrational}

For proving that $P\circ P \subseteq P$,
we adapted the programs of Section~\ref{upper1}, but the
process of computation was not so straightforward and ``automatic''
as we had hoped.
By construction, the vectors defining $P$ are irrational.
\iffull
As we will now discuss,
\else
As discussed in Appendix~\ref{irrational},
\fi
it is unavoidable to treat certain operations with these vectors as exact
operations.
\begin{figure}[htb]
  \centering
  \includegraphics{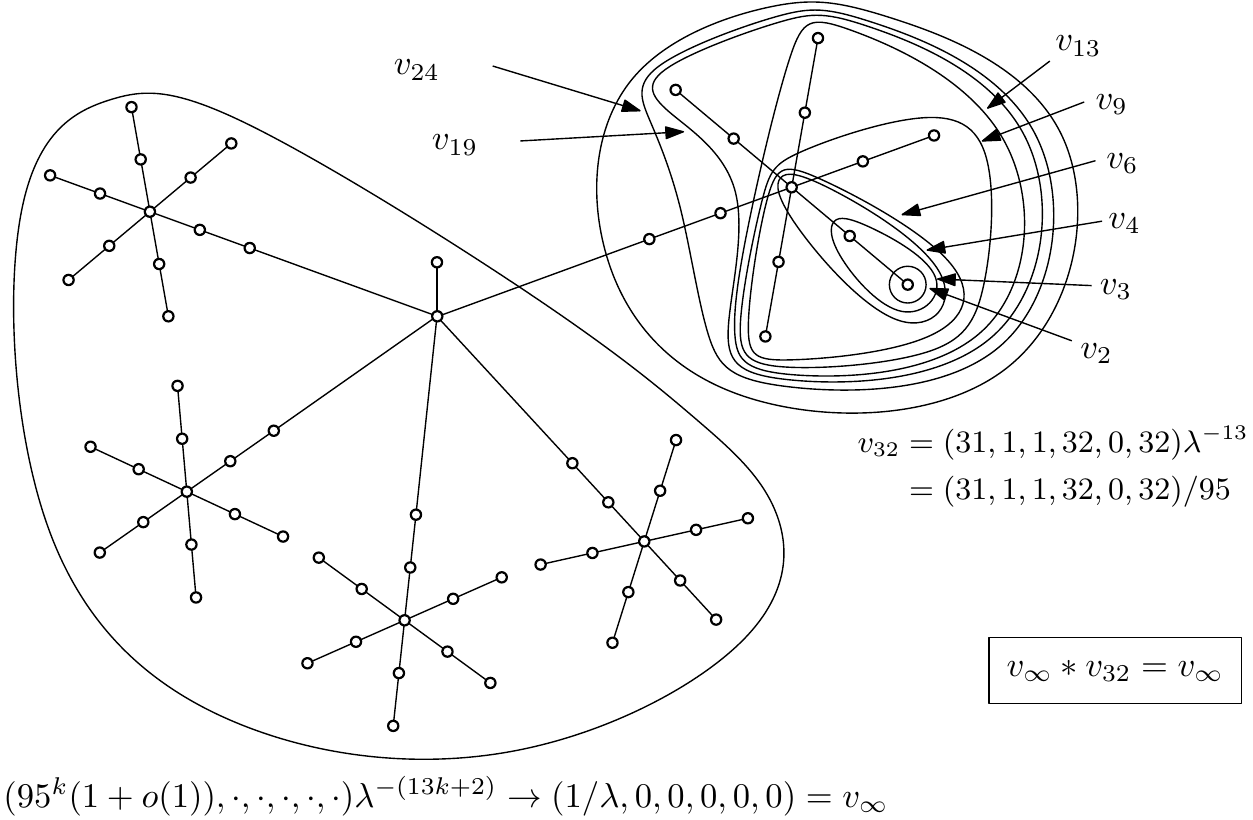}
  \caption{Adding another snowflake to
    a star of   $k\to\infty$
 snowflakes    }
  \label{fig:irrational}
\end{figure}

\long\def\irrationalvertices{%
As illustrated in Figure~\ref{fig:irrational},
there is a chain of $\star$ operations,
starting with the seed value
$v_{2}$, 
and leading via
$v_{3},
v_{6},
v_{9},
v_{13},
v_{19},
v_{24}$ to the vector 
$v_{32}=
(31, 1, 1, 32, 0, 32)/95$, which
 corresponds to the snowflake rooted at one of its leaves.
If these calculations were done imprecisely, then to maintain a
conservative approximation, $P$ would contain a value
 $\tilde v_{32}$ which is larger than the true value
 $v_{32}$ in all non-zero components.

We shall now argue that such a value cannot exist in a bounded set $P$
which is closed under the $\star $-operation.
The reason is the relation $v_1\star v_{32}=v_1$,
 which arises naturally
from the definition of the stars of snowflakes: Adding
another snowflake to a star of snowflakes yields a bigger star of
snowflakes.
In the limit, the relation expressing this composition converges to
 $v_\infty\star v_{32}=v_\infty$, and since $v_1$ is just a scaled copy of $v_\infty$, we
 also have
 $v_1\star v_{32}=v_1$.
 
Expressing this differently,
the linear
function
$v\mapsto v\star v_{32}$ has $v_1$ as an eigenvector with eigenvalue~1.
 With the modified value,
 $v_1\star \tilde v_{32}$ would be strictly larger than $v_1$ in the first
 component. Thus, the $\star $ operation with 
 $\tilde v_{32}$ acts on $v_1$ like a multiplication with a factor $F$
 strictly larger than 1.
The same holds true when $v_1$ is replaced by another non-zero vector of the
form $(x,0,0,0,0,0)$. By monotonicity, the first component of
\emph{any}
vector in $P$ (such as the vector  $\tilde v_{32}$ itself, for instance)
increases at least by
the factor $F$ when it 
 is multiplied by~$\tilde v_{32}$.
 It follows that $P$ cannot remain bounded.
}
\iffull
\irrationalvertices
\fi

 When constructing the set of vectors, we would have liked to use
 exact computation, but software that would perform exact linear
 programming with algebraic inputs was not readily available.  Thus we
 used standard floating-point linear-programming computations to prune
 points of $Q\circ Q$ in the interior of the convex hull, but as
\iffull
 we mentioned earlier,
\else
 mentioned in Appendix~\ref{Implementation},
\fi
this is not reliable.

\subsection{Certification of the results}
\label{sec:certify}

To turn this computation into a proof, we extracted from the
 linear-programming solutions the coefficients which certified that a
 point is majorized by a convex combination of other points.  We
 rounded these coefficients 
 to multiples of $0.0001$ while ensuring that their sum remains 1, and
 wrote them to a file.
For illustration,
we report
in
 Appendix~\ref{calculate9}
the certifying
coefficients for 
all
products $v_9\star v_j$\iftrue, $j=1,\ldots,55$\fi.

We then used a separate program to show that $v_i\star  v_j \in P$ for
all pairs of vertices $v_i,v_j$.
 The cases when the result is equal to
another vertex of $P$
are treated separately.
The complete list of these
cases is
in the left column of Table~\ref{tab:P}. 
These cases can be checked with integer arithmetic,
taking out common factors of $\lambda$. 
The only exception is the
equation $v_{1}\star v_{32}=v_1$, but this can also be checked by a simple
integer calculation since
$\lambda^{-13}=1/95$, and the fractional factor 0.9 is common on both
sides and is therefore irrelevant.

The remaining conditions were checked by floating-point calculations, using
the stored coefficients from the file. The smallest gap
occurred when showing that $v_{51}\star v_{41} \preceq v_{21}$. This
elementwise comparison holds by a
margin of $4.7\times 10^{-6}$, which is far bigger than the accuracy
of floating-point computations. The checking calculations involve only
additions and multiplications of positive numbers.
The largest power of $\lambda^{-1}$ that occurs is 54, for
computing $v_{55}\star v_{55}$, and there are just a couple of dozen more arithmetic steps
before the final comparison is made for each pair $i,j$.
Thus, errors do not accumulate over long sequences of calculations,
and even
single-precision floating-point calculations would safe to use for
checking this part of the
proof.
The checking program is about 130 lines of \textsc{Python} code,
including also the exact equality tests.
The file with data for the 55 vertices of $P$ (Table~\ref{tab:P})
has 1774 bytes, and the file with the coefficients of the $55^2$ inequalities
 certifying that $P\circ P \subseteq P$
 has 128 kBytes.
 \ifARXIV The three files
\texttt{minimal-dominating-sets-in-trees-docheck.py},
 \texttt{hullvertices.py}, and
 \texttt{lambdas.py}
 are contained in the source bundle of this arXiv preprint.
 \fi

By evaluating $\bar M$ for the vertices of $P$, one finds that the
maximum, $2/\lambda^2\approx 0.99257841$ is achieved by $v_3$,
corresponding to the tree with two vertices. This implies
$M_n \le 0.992579\lambda^n$,
thus proving part~1 of Theorem~\ref{sizes}.

\ifsoda
\else

To illustrate some of the difficulties that we faced when
trying to find a reliable proof, we finish this section with
the report of two failed calculation attempts with the use of
floating-point linear-programming software.

(i) As argued above, a natural point to consider as a vertex of $P$ is
the point $v_\infty=(1/\lambda,0,0,0,0,0)$. We started the calculation by putting with $v_\infty$ into
$Q$ instead of $v_1$, together with the vectors $v_{2}, v_{3}, v_{6},
v_{9}, v_{13}, v_{19}, v_{24}, v_{32}$, for which we know that they
must lie on the boundary of $P$.  The hull $Q$ stabilized with a set
of 89 vertices after a couple of minutes. However, when we
tried to check and
reproduce 
 the coefficients that
were extracted from the linear program with more accurate arithmetic,
we failed.
This setup should lead to the ``correct'' hull $P=\hullp(P_0)$.
However, 
 we do not even know whether this set (or rather, its topological closure)
 is at all a polytope with finitely many
vertices,
or whether this approach is doomed unless one adds artificial
points like our point $v_1$.

(ii) For comparison, we omitted both vectors $v_\infty$ and $v_1$
altogether.
For this case, we \emph{know} theoretically that $P$ should grow closer and closer
to $v_\infty$ but should never reach it.
However, even in this case, the program
terminated after a few minutes, with a hull of 94 vertices.

\fi
\section{Outlook and Open Questions}



\subsection{The growth of a bilinear operation}
\label{sec:bilinear}

We have already mentioned
in Section~\ref{characteristic-vectors}
that the bilinear operation $\star$ on
sextuples captures all the necessary information of the counting
question, together with the starting vector $v_0$ and the terminal function
$\bar M$ from~\eqref{eq:total}.
Once we know these algebraic data, we
can
abstract from the
background of the original \mds s problem:
%
%
What is the largest value that can be built by combining
$n$ copies of $v_0$ with $n-1$ applications of the
(non-associative) operation $\star$, and how fast does this value grow
with~$n$?
For example, with $n=9$ elements,
we could build the expression
$$
\bar M 
((v_0\star (v_0\star ((v_0\star v_0)\star (v_0\star (v_0\star v_0)))))\star (v_0\star
v_0)).
$$

When we ask the analogous question for a \emph{linear} operation
$f\colon \reals^d\to \reals^d$, this is a basic problem of linear
algebra that is well-understood. The answer is given by the dominant
eigenvalue of $f$, and the growth does not depend on the starting
vector (except for degenerate cases).  What happens for a
general
bilinear operation $\star\colon \reals^d\times\reals^d\to\reals^d$?
This question is open for further study.
Let us assume that the operation has nonnegative coefficients.
Proposition~\ref{characterize} gives a characterization of the
exponential growth rate in terms of a convex body~$P$. Is it
sufficient to consider bodies $P$ that are polytopes?
With the correct choice of $\lambda$, will the iterative process
 converge to a polytope?
How does the growth depend on
the starting vector?  When is there a single ``characteristic'' body $P$ that works for
all starting vectors?
If the growth rate always attained by a ``periodic'' constructions, like our star of snowflakes?
Is the growth rate necessarily an algebraic number?
Is it computable or approximable?

The following speculative
argument tries to explain why it might be no coincidence that 
$\lambda$ turned out to be algebraic for \mds s.
Perhaps these thoughts can be strengthened generalized to show that the growth
rate is always an algebraic number.
In
our polytope
$P$ that we used for proving the upper bound of
Theorem~\ref{sizes} (Table~\ref{tab:P}),
a typical vertex $v$ has 
an implicit power $v=\lambda^iu$ according to
how it is generated, telling how it varies in terms of $\lambda$.
The tight case, when $\lambda$ cannot be improved
without violating the condition $P\circ P \subseteq P$,
 is characterized
by some point $\lambda^i u$ lying on the boundary of $P$, i.e.,
 in some hyperplane through some vertices $\lambda^{i_k} u_k$.
 This condition generates a polynomial equation in $\lambda$,
 and thus, 
 $\lambda$ is an algebraic number.
 (In our case, the critical equation is
 $v_{1}\star v_{32}=v_1$ as explained in Section~\ref{sec:irrational}.
 Since $v_1$ was not chosen in the form $v=\lambda^iu$, the above
 argument is not strictly valid in this case.)


In the case of linear operators, the growth is determined by the
eigenvalues.  Eigenvalues have been considered also for bilinear (and
multilinear) operations, but the usual approach it to set up an eigenvector
equation of the form $x\star x = \lambda x$ (as it would be written in
our notation) and investigate the solutions and the algebraic
properties of this system, see for example \cite{tensors,breiding}.
Are the eigenvectors and eigenvalues in this sense related to the
growth rate for our question?

%
Finally, it is interesting to note that some
problem-specific
properties that we see in trees can be written as 
algebraic properties of the $\star$-operation.
We list a few of
them. 


%

\begin{itemize}
\item 
It is clear that the order in which subtrees are added is
irrelevant. This is reflected in the following
 ``right commutative law'':
 \begin{displaymath}
(u
\star 
v)
\star 
w
=(u
\star 
w)
\star 
v
\end{displaymath}
\item 
At the level of counting \mds s, it does not matter which node is
chosen as the root. This is reflected in the following partial
 commutativity law under the operator~$\bar M$:
\[ \bar M(u\star v)=\bar M(v\star u) \]
\item 
 Observation~\ref{obse}(\ref{twins}) says that
twins are irrelevant as far as \mds s are concerned:
\[(v \star  v_0)\star  v_0= v\star  v_0\] 

\item One property that cannot be directly expressed in purely algebraic
terms is the supermultiplicativity of $M_n$.
But the main case of its proof,
Observation~\ref{obse}(\ref{super-trees}), can be reduced to a pure
calculation:
It says that
the combination of two trees where each root has
a leaf as a neighbor
will multiply the
number of solutions of the two subtrees:
$$\bar M((v \star  v_0)\star (w\star  v_0)) = \bar M(v \star  v_0)\cdot\bar M(w\star  v_0)$$
This holds even
in a stronger form
than needed,
as the  vector equation
$$(v \star  v_0)\star (w\star  v_0) = v \star  v_0\cdot\bar M(w\star  v_0).$$

\end{itemize}

All these equations
can be
 checked computationally
by substituting the definitions and expanding the terms,
preferable
with a computer
algebra system.

\subsection{Other applications of the method}
\label{sec:conc}

Proposition~\ref{characterize} and the algorithm of
Section~\ref{automatic}
give a versatile method for investigating growth problems that come
from
dynamic-programming recursions.
This 
extends beyond trees to other
structures that can be hierarchically built up in a tree-like fashion.
As a next step, one might consider 2-trees or  series-parallel graphs.
 The combinatorial case analysis leading to the ``$\star$'' operations will be more complicated.
For example, for series-parallel
graphs, one has to monitor the status of \emph{two} terminal vertices
instead of just one root vertex, and
the number of categories will multiply.

In Section~\ref{sec:optimize}, we were interested in the \emph{minimum} number
of \mds s in \emph{trees
  without twins}. Here the method
of Proposition~\ref{characterize}
has to be adapted. We have to maintain two sets
of sixtuples, distinguishing whether the root has a leaf neighbor or not.

One can also count other structures than \mds s, for example
\emph{maximal irredundant subsets} of vertices.
In an \emph{irredundant} set, 
every vertex
has a private neighbor, but the set does not have to be dominating.


\iffull

\subsection{Loopless enumeration and Gray codes}
\label{sec:gray}

In Section~\ref{sec:optimize}, we discussed the possibility to
generate \mds s $D$ faster than in linear time per solution, by counting
only the operations to insert or remove an element from $D$.
A~more ambitious goal would be to enumerate the solutions
with \emph{constant
delay}. Such enumeration algorithms are called \emph{loopless} or \emph{loop-free}, see for example~\cite{Ehrlich,kn4,hr-lgcet-18}.
The sequence in which the solutions are generated has to have the
property that the difference between
consecutive solutions is bounded in size by a constant. Such a sequence
is called a \emph{Gray code}, in analogy with the classical Gray code that
goes through
all 0-1-sequences of a given length by flipping single bits at a time.


We have already seen in Figure~\ref{fig:stars}
in Section~\ref{sec:optimize}
that a Gray code is
impossible without preprocessing, and we have argued
that it makes sense to restrict our
attention to trees without twins.
Is there a Gray code through all \mds s for this class of trees?
To define such a Gray code in an inductive way, one might look at 
Table~\ref{tab:combine},
remembering its interpretation as an equation for sets, and navigate the
table in a clever way.




\fi

\ifTALG
\begin{acks}
  This work was initiated at the Lorentz Center workshop on
  ``Enumeration Algorithms Using Structure'' in Leiden, the
  Netherlands, August 24--28, 2015.
\end{acks}
\else  
\paragraph{Acknowledgment.}
  This work was initiated at the Lorentz Center workshop on
  ``Enumeration Algorithms Using Structure'' in Leiden, the
  Netherlands, August 24--28, 2015.
\fi

\ifTALG
\bibliographystyle{ACM-Reference-Format}
\newcommand{\urlWOheader}{}
\def\urldef\tempurl\url#1{\unskip\unskip\unskip\unskip\ \urlWOheader #1\pxr}
\def\urlWOheader http#1#2#3#4#5/#6\pxr{{%
    \def\first{#1#2#3#4#5}%
    \def\doistart{s://doi.org}%
    \ifx\first\doistart
      \gdef\tempurl{\texttt{DOI:}\penalty0\href{http#1#2#3#4#5/#6}{#6}}%
    \else
      \gdef\tempurl{\url{http#1#2#3#4#5/#6}}%
    \fi
  }}

\else
\bibliographystyle{abbrv}
\fi
\goodbreak
\bibliography{mindom}


\begin{thebibliography}{00}


\ifx \showCODEN    \undefined \def \showCODEN     #1{\unskip}     \fi
\ifx \showDOI      \undefined \def \showDOI       #1{{\tt DOI:}\penalty0{#1}\ }
  \fi
\ifx \showISBNx    \undefined \def \showISBNx     #1{\unskip}     \fi
\ifx \showISBNxiii \undefined \def \showISBNxiii  #1{\unskip}     \fi
\ifx \showISSN     \undefined \def \showISSN      #1{\unskip}     \fi
\ifx \showLCCN     \undefined \def \showLCCN      #1{\unskip}     \fi
\ifx \shownote     \undefined \def \shownote      #1{#1}          \fi
\ifx \showarticletitle \undefined \def \showarticletitle #1{#1}   \fi
\ifx \showURL      \undefined \def \showURL       #1{#1}          \fi
\providecommand\bibfield[2]{#2}
\providecommand\bibinfo[2]{#2}
\providecommand\natexlab[1]{#1}
\providecommand\showeprint[2][]{arXiv:#2}

\bibitem[\protect\citeauthoryear{Breiding}{Breiding}{2017}]%
        {breiding}
\bibfield{author}{\bibinfo{person}{Paul Breiding}.}
  \bibinfo{year}{2017}\natexlab{}.
\newblock {\em \bibinfo{title}{Numerical and Statistical Aspects of Tensor
  Decompositions}}.
\newblock \bibinfo{thesistype}{Ph.D. Dissertation}. \bibinfo{school}{Technische
  Universit\"at Berlin}.
\newblock
\showDOI{%
\url{http://dx.doi.org/10.14279/depositonce-6148}}


\bibitem[\protect\citeauthoryear{Couturier, Heggernes, van~'t Hof, and
  Kratsch}{Couturier et~al\mbox{.}}{2013}]%
        {COUTURIER201382}
\bibfield{author}{\bibinfo{person}{Jean-Fran\c{c}ois Couturier},
  \bibinfo{person}{Pinar Heggernes}, \bibinfo{person}{Pim van~'t Hof}, {and}
  \bibinfo{person}{Dieter Kratsch}.} \bibinfo{year}{2013}\natexlab{}.
\newblock \showarticletitle{Minimal dominating sets in graph classes:
  Combinatorial bounds and enumeration}.
\newblock \bibinfo{journal}{{\em Theoretical Computer Science\/}}
  \bibinfo{volume}{487} (\bibinfo{year}{2013}), \bibinfo{pages}{82--94}.
\newblock
\showISSN{0304-3975}
\showDOI{%
\url{http://dx.doi.org/10.1016/j.tcs.2013.03.026}}


\bibitem[\protect\citeauthoryear{Ehrlich}{Ehrlich}{1973}]%
        {Ehrlich}
\bibfield{author}{\bibinfo{person}{Gideon Ehrlich}.}
  \bibinfo{year}{1973}\natexlab{}.
\newblock \showarticletitle{Loopless Algorithms for Generating Permutations,
  Combinations, and Other Combinatorial Configurations}.
\newblock \bibinfo{journal}{{\em J. Assoc. Comput. Mach.\/}}
  \bibinfo{volume}{20}, \bibinfo{number}{3} (\bibinfo{date}{July}
  \bibinfo{year}{1973}), \bibinfo{pages}{500--513}.
\newblock
\showISSN{0004-5411}
\showDOI{%
\url{http://dx.doi.org/10.1145/321765.321781}}


\bibitem[\protect\citeauthoryear{Golovach, Heggernes, Kant{\'e}, Kratsch, and
  Villanger}{Golovach et~al\mbox{.}}{2017}]%
        {ghkkv-mdsig-17}
\bibfield{author}{\bibinfo{person}{Petr Golovach}, \bibinfo{person}{Pinar
  Heggernes}, \bibinfo{person}{Mamadou~Moustapha Kant{\'e}},
  \bibinfo{person}{Dieter Kratsch}, {and} \bibinfo{person}{Yngve Villanger}.}
  \bibinfo{year}{2017}\natexlab{}.
\newblock \showarticletitle{{Minimal dominating sets in interval graphs and
  trees}}.
\newblock \bibinfo{journal}{{\em Discrete Applied Mathematics\/}}
  \bibinfo{volume}{216} (\bibinfo{date}{Jan.} \bibinfo{year}{2017}),
  \bibinfo{pages}{162--170}.
\newblock
\showDOI{%
\url{http://dx.doi.org/10.1016/j.dam.2016.01.038}}


\bibitem[\protect\citeauthoryear{Herter and Rote}{Herter and Rote}{2018}]%
        {hr-lgcet-18}
\bibfield{author}{\bibinfo{person}{Felix Herter} {and}
  \bibinfo{person}{G{\"u}nter Rote}.} \bibinfo{year}{2018}\natexlab{}.
\newblock \showarticletitle{Loopless {Gray} code enumeration and the {Tower} of
  {Bucharest}}.
\newblock \bibinfo{journal}{{\em Theoretical Computer Science\/}}
  \bibinfo{volume}{748} (\bibinfo{year}{2018}), \bibinfo{pages}{40--54}.
\newblock
\showDOI{%
\url{http://dx.doi.org/10.1016/j.tcs.2017.11.017}}


\bibitem[\protect\citeauthoryear{Knuth}{Knuth}{2011}]%
        {kn4}
\bibfield{author}{\bibinfo{person}{Donald~E. Knuth}.}
  \bibinfo{year}{2011}\natexlab{}.
\newblock \bibinfo{booktitle}{{\em Combinatorial Algorithms, Part~1}}.
  \bibinfo{series}{The Art of Computer Programming}, Vol.~\bibinfo{volume}{4A}.
\newblock \bibinfo{publisher}{Addison-Wesley}.
\newblock


\bibitem[\protect\citeauthoryear{Krzywkowski}{Krzywkowski}{2013}]%
        {k-tmmds-13}
\bibfield{author}{\bibinfo{person}{Marcin Krzywkowski}.}
  \bibinfo{year}{2013}\natexlab{}.
\newblock \showarticletitle{Trees having many minimal dominating sets}.
\newblock \bibinfo{journal}{{\em Inf. Process. Lett.\/}} \bibinfo{volume}{113},
  \bibinfo{number}{8} (\bibinfo{date}{April} \bibinfo{year}{2013}),
  \bibinfo{pages}{276--279}.
\newblock
\showISSN{0020-0190}
\showDOI{%
\url{http://dx.doi.org/10.1016/j.ipl.2013.01.020}}


\bibitem[\protect\citeauthoryear{Kungching, Liqun, and Tan}{Kungching
  et~al\mbox{.}}{2013}]%
        {tensors}
\bibfield{author}{\bibinfo{person}{Chang Kungching}, \bibinfo{person}{Qi
  Liqun}, {and} \bibinfo{person}{Zhang Tan}.} \bibinfo{year}{2013}\natexlab{}.
\newblock \showarticletitle{A survey on the spectral theory of nonnegative
  tensors}.
\newblock \bibinfo{journal}{{\em Numerical Linear Algebra with Applications\/}}
  \bibinfo{volume}{20}, \bibinfo{number}{6} (\bibinfo{year}{2013}),
  \bibinfo{pages}{891--912}.
\newblock
\showDOI{%
\url{http://dx.doi.org/10.1002/nla.1902}}


\bibitem[\protect\citeauthoryear{Rote}{Rote}{2019}]%
        {r-mnmds-19}
\bibfield{author}{\bibinfo{person}{G{\"u}nter Rote}.}
  \bibinfo{year}{2019}\natexlab{}.
\newblock \showarticletitle{The maximum number of minimal dominating sets in a
  tree}. In \bibinfo{booktitle}{{\em Proceedings of the 30th Annual ACM-SIAM
  Symposium on Discrete Algorithms (SODA19), San Diego}},
  \bibfield{editor}{\bibinfo{person}{Timothy Chan}} (Ed.).
  \bibinfo{publisher}{SIAM}, \bibinfo{pages}{1201--1214}.
\newblock
\showDOI{%
\url{http://dx.doi.org/10.1137/1.9781611975482.73}}


\end{thebibliography}

\appendix

\section{Certifying Computations for $v_9\star v_j$}
\label{calculate9}
For illustration, we show a section of the data that are used in the
proof of the closure property \eqref{closed} of the polytope $P$ in
Section~\ref{sec:certify}.
Such data exist for each product
$v_i\star v_j$, $1\le i,j\le55$.
The coefficients stand for exact four-digit decimal numbers, which add up to
1 on each line.
\smallskip

\ifjustsyntaxchecking\else
{\iffalse\else\small\fi
  \ifsoda\else
  \leftskip=\parindent
  \fi
{%
\let\prec=\preceq
\def\b{\hphantom0}
\noindent
$v_{9}\star v_{1\b} \prec v_{13}$\\
$v_{9}\star v_{2\b} \prec v_{3}$\\
$v_{9}\star v_{3\b} =v_{12}$\\
$v_{9}\star v_{4\b} \prec 0.0158\,v_{1\b}+0.1798\,v_{5\b}+0.4284\,v_{3\b}+0.3760\,v_{13}$\\
$v_{9}\star v_{5\b} =v_{19}$\\
$v_{9}\star v_{6\b} \prec v_{10}$\\
$v_{9}\star v_{7\b} \prec 0.1454\,v_{1\b}+0.1066\,v_{5\b}+0.2046\,v_{3\b}+0.5434\,v_{13}$\\
$v_{9}\star v_{8\b} =v_{27}$\\
$v_{9}\star v_{9\b} \prec v_{10}$\\
$v_{9}\star v_{10} \prec 0.0823\,v_{1\b}+0.1504\,v_{5\b}+0.0609\,v_{3\b}+0.7064\,v_{13}$\\
$v_{9}\star v_{11} \prec 0.2096\,v_{1\b}+0.0732\,v_{5\b}+0.0922\,v_{3\b}+0.6250\,v_{13}$\\
$v_{9}\star v_{12} =v_{37}$\\
$v_{9}\star v_{13} \prec v_{10}$\\
$v_{9}\star v_{14} \prec 0.0518\,v_{1\b}+0.1396\,v_{5\b}+0.7853\,v_{13}+0.0233\,v_{8\b}$\\
$v_{9}\star v_{15} \prec 0.1127\,v_{1\b}+0.0979\,v_{5\b}+0.0110\,v_{3\b}+0.7784\,v_{13}$\\
$v_{9}\star v_{16} \prec 0.2412\,v_{1\b}+0.0595\,v_{5\b}+0.0351\,v_{3\b}+0.6642\,v_{13}$\\
$v_{9}\star v_{17} =v_{43}$\\
$v_{9}\star v_{18} \prec 0.0171\,v_{20}+0.2959\,v_{3\b}+0.3547\,v_{28}+0.0909\,v_{5\b}+0.2414\,v_{27}$\\
$v_{9}\star v_{19} \prec v_{10}$\\
$v_{9}\star v_{20} \prec 0.0373\,v_{1\b}+0.0420\,v_{5\b}+0.8224\,v_{13}+0.0983\,v_{8\b}$\\
$v_{9}\star v_{21} \prec 0.0631\,v_{1\b}+0.0422\,v_{9\b}+0.0367\,v_{5\b}+0.8125\,v_{13}+0.0455\,v_{8\b}$\\
$v_{9}\star v_{22} \prec 0.1238\,v_{1\b}+0.0795\,v_{9\b}+0.0593\,v_{5\b}+0.7373\,v_{13}+0.0001\,v_{8\b}$\\
$v_{9}\star v_{23} \prec 0.2567\,v_{1\b}+0.0556\,v_{5\b}+0.0054\,v_{3\b}+0.6823\,v_{13}$\\
$v_{9}\star v_{24} =v_{48}$\\
$v_{9}\star v_{25} \prec 0.0911\,v_{5\b}+0.3149\,v_{3\b}+0.3847\,v_{28}+0.1192\,v_{37}+0.0901\,v_{27}$\\
$v_{9}\star v_{26} \prec v_{10}$\\
$v_{9}\star v_{27} \prec 0.0310\,v_{1\b}+0.0150\,v_{12}+0.8388\,v_{13}+0.1152\,v_{8\b}$\\
$v_{9}\star v_{28} \prec 0.0393\,v_{1\b}+0.0632\,v_{9\b}+0.0222\,v_{12}+0.8270\,v_{13}+0.0483\,v_{8\b}$\\
$v_{9}\star v_{29} \prec 0.0650\,v_{1\b}+0.1454\,v_{9\b}+0.0157\,v_{5\b}+0.7462\,v_{13}+0.0277\,v_{8\b}$\\
$v_{9}\star v_{30} \prec 0.1296\,v_{1\b}+0.1188\,v_{9\b}+0.0226\,v_{5\b}+0.7151\,v_{13}+0.0139\,v_{8\b}$\\
$v_{9}\star v_{31} \prec 0.0654\,v_{1\b}+0.1382\,v_{9\b}+0.2283\,v_{3\b}+0.0405\,v_{5\b}+0.5276\,v_{13}$\\
$v_{9}\star v_{32} \prec 0.2643\,v_{1\b}+0.0243\,v_{5\b}+0.6898\,v_{13}+0.0216\,v_{8\b}$\\
$v_{9}\star v_{33} \prec 0.9543\,v_{49}+0.0188\,v_{3\b}+0.0269\,v_{45}$\\
$v_{9}\star v_{34} \prec 0.0942\,v_{5\b}+0.3247\,v_{3\b}+0.3909\,v_{28}+0.1772\,v_{37}+0.0130\,v_{27}$\\
$v_{9}\star v_{35} \prec v_{10}$\\
$v_{9}\star v_{36} \prec 0.0284\,v_{1\b}+0.0667\,v_{12}+0.8450\,v_{13}+0.0599\,v_{8\b}$\\
$v_{9}\star v_{37} \prec 0.0277\,v_{1\b}+0.0813\,v_{9\b}+0.0661\,v_{12}+0.8249\,v_{13}$\\
$v_{9}\star v_{38} \prec 0.0365\,v_{1\b}+0.1762\,v_{9\b}+0.0151\,v_{12}+0.7502\,v_{13}+0.0220\,v_{8\b}$\\
$v_{9}\star v_{39} \prec 0.0665\,v_{1\b}+0.1946\,v_{9\b}+0.0036\,v_{5\b}+0.7132\,v_{13}+0.0221\,v_{8\b}$\\
$v_{9}\star v_{40} \prec 0.1329\,v_{1\b}+0.1365\,v_{9\b}+0.0028\,v_{5\b}+0.7039\,v_{13}+0.0239\,v_{8\b}$\\
$v_{9}\star v_{41} \prec 0.0279\,v_{1\b}+0.0944\,v_{12}+0.8461\,v_{13}+0.0316\,v_{8\b}$\\
$v_{9}\star v_{42} \prec 0.0210\,v_{1\b}+0.1225\,v_{9\b}+0.0652\,v_{12}+0.7913\,v_{13}$\\
$v_{9}\star v_{43} \prec 0.0229\,v_{1\b}+0.1975\,v_{9\b}+0.0354\,v_{12}+0.7442\,v_{13}$\\
$v_{9}\star v_{44} \prec 0.0359\,v_{1\b}+0.2301\,v_{9\b}+0.0148\,v_{12}+0.7119\,v_{13}+0.0073\,v_{8\b}$\\
$v_{9}\star v_{45} \prec 0.0679\,v_{1\b}+0.2169\,v_{9\b}+0.0080\,v_{12}+0.6966\,v_{13}+0.0106\,v_{8\b}$\\
$v_{9}\star v_{46} \prec 0.0185\,v_{1\b}+0.1434\,v_{9\b}+0.0661\,v_{12}+0.7720\,v_{13}$\\
$v_{9}\star v_{47} \prec 0.0162\,v_{1\b}+0.2226\,v_{9\b}+0.0360\,v_{12}+0.7252\,v_{13}$\\
$v_{9}\star v_{48} \prec 0.0213\,v_{1\b}+0.2550\,v_{9\b}+0.0218\,v_{12}+0.7019\,v_{13}$\\
$v_{9}\star v_{49} \prec 0.0363\,v_{1\b}+0.2568\,v_{9\b}+0.0163\,v_{12}+0.6906\,v_{13}$\\
$v_{9}\star v_{50} \prec 0.0135\,v_{1\b}+0.2352\,v_{9\b}+0.0379\,v_{12}+0.7134\,v_{13}$\\
$v_{9}\star v_{51} \prec 0.0144\,v_{1\b}+0.2721\,v_{9\b}+0.0232\,v_{12}+0.6903\,v_{13}$\\
$v_{9}\star v_{52} \prec 0.0211\,v_{1\b}+0.2834\,v_{9\b}+0.0167\,v_{12}+0.6788\,v_{13}$\\
$v_{9}\star v_{53} \prec 0.3716\,v_{20}+0.3132\,v_{28}+0.2973\,v_{21}+0.0179\,v_{24}$\\
$v_{9}\star v_{54} \prec 0.0144\,v_{1\b}+0.2965\,v_{9\b}+0.0184\,v_{12}+0.6707\,v_{13}$\\
$v_{9}\star v_{55} \prec 0.3078\,v_{20}+0.3709\,v_{28}+0.3010\,v_{21}+0.0203\,v_{24}$
}

}
\fi

\iffull
\section{Another Enumeration Algorithm: ENUM3}
\label{variation-enum}

We present another variation of an algorithm for enumerating
 \mds s through the expression DAG.
It combines the positive features of Algorithms ENUM1 and ENUM2.
In the outer loop of product nodes, subtrees where nothing
changes are not visited,
potentially saving a lot of work.
In this respect, we follow ENUM1.
Like ENUM2, the end of a loop is signaled simultaneously with the
delivery of the last solution.
Thus, the dummy visits of  ENUM1 are avoided.
Unlike ENUM2, we also distinguish the first element of a loop with a
special message.




\ifjustsyntaxchecking\else
\begin{figure}

  \begin{tabbing}
    \hskip 3em\=\+
\quad\= \hskip2,7em\= \hskip2,7em\=\hskip2,7em\=\hskip2,7em\=\hskip2,7em\=\hskip2,7em\=\kill
Let $K$ be the master node.\\
\mes\ := PRODUCE-FIRST,
go to the target node, and start the following loop.\\[0.6ex]
\textbf{loop}\+\\
let $K$ be the current node
\\
  \textbf{case} $K$ is a basis node for vertex $a$:\+\\
\textbf{case} $K$ represents the set $\{a\}$:\\
\>insert vertex $a$ into $D$ if it is not already in $D$\\
\textbf{case} $K$ represents the set $\emptyset$:\\
\>remove vertex $a$ from $D$ if it is in $D$\\
\mes\ := LAST, 
and go to the parent\-\\
  \textbf{case} $K$ is the master node:\+\\
  report the current solution $D$\\
\textbf{case} \mes\  = DONE:\+\\
\mes\  := PRODUCE-NEXT, 
and go to the target node\-\\
\textbf{case} \mes\  = LAST:\+\\
exit from the loop and stop\-\-\\
%
%
  \textbf{case} $K$ is a union node:\+\\
\textbf{case} \mes\  = PRODUCE-FIRST:\+\\
$K.\sta :=$ ``child 1''\\
\mes\  := PRODUCE-FIRST, 
and go to the first child\-\\
\textbf{case} \mes\  = PRODUCE-NEXT:\+\\
\textbf{case} $K.\sta =$ ``child 1'':\+\\
\mes\  := PRODUCE-NEXT, 
and go to the first child\-\\
\textbf{case} $K.\sta = {}$``transition from child 1 to child 2'':\+\\
$K.\sta :=$ ``child 2''\\
\mes\  := PRODUCE-FIRST, 
and go to the second child\-\\
\textbf{case} $K.\sta =$ ``child 2'':\+\\
\mes\  := PRODUCE-NEXT, 
and go to the second child\-\-\\
\textbf{case} \mes\  = DONE:\+\\
\mes\  := DONE, 
and go to the parent\-\\
\textbf{case} \mes\  = LAST:\+\\
\textbf{case} $K.\sta ={}$``child 1'':\+\\
$K.\sta := {}$``transition from child 1 to child 2''\\
\mes\  := DONE, 
and go to the parent\-\\
\textbf{case} $K.\sta =$ ``child 2'':\+\\
$K.\sta := {}$``dormant''\\
\mes\  := LAST, 
and go to the parent\-\-\-\\
  \textbf{case} $K$ is a product node:\+\\
handle $K$ by the algorithm in Figure~\ref{fig:product-algorithm}
\end{tabbing}
  \caption{Algorithm ENUM3}
  \label{fig:ENUM3-algorithm}
\end{figure}

\begin{figure}
  
  \begin{tabbing}
        \hskip 3em\=\+ 
\quad\= \hskip2,7em\= \hskip2,7em\=\hskip2,7em\=\hskip2,7em\=\hskip2,7em\=\hskip2,7em\=\kill
\>\+
  \textbf{case} $K$ is a product node:\+\\
\textbf{case} \mes\  = PRODUCE-FIRST:\+\\
$K.\sta := {}$``working''\\
$K.\child := 1$\\
\mes\  := PRODUCE-FIRST, 
and go to the first child\-\\
\textbf{case} \mes\  = PRODUCE-NEXT:\+\\
\textbf{case} $K.\sta = {}$``working'' \textbf{or}  $K.\sta = {}$``child 1 has finished'':\+\\
$K.\child := 2$\\
\mes\  := PRODUCE-NEXT, 
and go to the second child\-\\
\textbf{case} $K.\sta = {}$``child 2 has finished'':\+\\
$K.\sta := {}$``working''\\
$K.\child := 1$\\
\mes\  := PRODUCE-NEXT, 
and go to the first child\-\-\\
\textbf{case} $K.\child = 1$ \textbf{and} \mes\  = DONE:\+\\
$K.\child := 2$\\
\mes\  := PRODUCE-FIRST, 
and go to the second child\-\\
\textbf{case} $K.\child = 1$ \textbf{and} \mes\  = LAST:\+\\
$K.\sta :=$ ``child 1 has finished''\\
$K.\child := 2$\\
\mes\  := PRODUCE-FIRST, 
and go to the second child\-\\
\textbf{case} $K.\child = 2$ \textbf{and} \mes\  = DONE:\+\\
\mes\  := DONE, 
and go to the parent\-\\
\textbf{case} $K.\child = 2$ \textbf{and} \mes\  = LAST:\+\\
\textbf{case}
$K.\sta =$ ``child 1 has finished'':\+\\
$K.\sta :=$ ``dormant''\\
\mes\  := LAST, 
and go to the parent\-\\
\textbf{case} $K.\sta =$ ``working'':\+\\
$K.\sta :=$ ``child 2 has finished''\\
\mes\  := DONE, 
and go to the parent\-
\end{tabbing}
  \caption{Algorithm ENUM3: Handling of a product node}
  \label{fig:product-algorithm}
\end{figure}

\fi

The algorithm is shown in Figures~\ref{fig:ENUM3-algorithm} and~\ref{fig:product-algorithm}.
Like Algorithm ENUM2 in  Section~\ref{enumeration}, this is a
low-level description without generators or coroutines. All message
passing is explicit.
However,
the algorithm is presented in a different style from ENUM2:
Instead of a family of patterns like Figures~\ref{fig:product}--\ref{fig:master},
the algorithm
is written more conventionally as a series of nested case distinctions.
Certain operations that
have been left out in Section~\ref{enumeration} are explicitly
stated, for example,
remembering the child of a product node that is currently visited (or
recognizing it when a message is received from it).
This changed style reflects the author's insecurity 
about the best way 
to present such enumeration
algorithms.

We shall now discuss some details.
 Messages are sent across the arcs of the expression DAG.
%
There are two types of \emph{request messages}:
PRODUCE-FIRST and PRODUCE-NEXT. They always flow downward in the
network, from the root towards the leaves.
There are two types of \emph{reply messages}:
DONE and LAST.
They always flow upward in the network.

Every union and product node has a \sta\ attribute from a small choice of possibilities.
In addition, every product node
records which of
its children has received a message
 in its \child\ attribute.
As in the algorithms of Section~\ref{listing}, we have an
additional \emph{master node} with a single outgoing arc to the target
node. Its only job is to send PRODUCE-NEXT requests until it receives
a LAST message that signals completion of the enumeration.

The current node 
is denoted by a
global variable $K$.  Depending on the type of node and on the message
received, the program may consult the \child\ or \sta\ attributes of the current
node. It will then possibly update the
attributes, and move to
an adjacent node with a new message, which is stored in the global
variable \mes.
The solution
 $D$ is maintained as another global variable.

 As in Algorithm ENUM2 in Section~\ref{enum2},
we explore various subtrees of the expression DAG in a
depth-first search manner, and
 we maintain a ``call stack'' of
 nodes that are still expecting a reply.
In the program, 
``go to node $K'$'' means:
push the current node $K$ on the stack, and set $K := K'$, while
``go to the parent'' means:
pop $K$ from the stack.


 The algorithm carries out very simple
operations, but it is not apparent what happens.
We will discover some structure by 
 describing the process from multiple views: from a single arc and
then from a single node.

\goodbreak
\paragraph{Message flow along an arc.}

The flow of messages along an arc is a strict alternation:
\nobreak
\begin{quote}
$\to$ request(PRODUCE-FIRST)\\
$\from$ 
reply(DONE)\\
$\to$ request(PRODUCE-NEXT)\\
$\from$ reply(DONE)\\
\dots\\
$\to$ request(PRODUCE-NEXT)\\
$\from$ reply(LAST)  
\end{quote}
A reply message signals that a
solution has been set up in
the vertices of the subtree associated to the child.
If no more solutions are available after the current one, this is signaled by
the LAST reply.

Since we have ensured that every node represents a {nonempty} set
of solutions, the PRODUCE-FIRST request will always produce a reply.
Thus,
 the minimum
total number of messages is two.
After a block is finished
with a LAST reply, a new block of messages can be initiated
with another PRODUCE-FIRST message.

In contrast to the algorithm ENUM2 of Section~\ref{enumeration},
there is a special 
PRODUCE-FIRST request
to initiate the dialogue.
This allows the node to know when it
needs to initialize itself. It also has the nice feature that it
 makes the message exchange pattern symmetric with
respect to the reversal of time.

When we now analyse the flow from the point of view of the different
types of nodes, we will inductively assume that the message exchange with the
children (if any) follows the pattern described above, and we will
follow the operation of the node from the initial
PRODUCE-FIRST request received from the parent to the final LAST reply.
The \sta\  of all union and product nodes is
initialized to
``dormant'', indicating that they are ready to
receive a
PRODUCE-FIRST message and start producing results.
 The
 ``dormant'' state has
actually
 only informational value without effect for the
algorithm.

\paragraph{Basis nodes.}
The basis nodes return immediately with a LAST message after
setting up the solution $D$
by inserting a vertex into $D$ or removing
it from~$D$.


\begin{figure}[b]
  \ifjustsyntaxchecking\else
  \centering
  
{
  \lineskiplimit=-1,6 \baselineskip
  \baselineskip = 0,55 \baselineskip
  \def\ler{\noalign{\hrule height-2pt}}

  \noindent \vbox{
\halign{\strut
  \hfil# & \hfil# \hfil & # \hfil&\qquad # \hfil\cr
message from/to parent & { child } &message from/to child&\sta\cr
\noalign{\hrule}
&&&dormant\cr
 PRODUCE-FIRST $\to$ &  1   & $\to$ PRODUCE-FIRST \cr
&&&  child 1\cr
DONE $\from$  &  1   & $\from$ DONE \cr
&&&  child 1\cr
PRODUCE-NEXT $\to$ &  1   & $\to$ PRODUCE-NEXT \cr
&&&  child 1\cr
DONE $\from$  &  1   & $\from$ DONE \cr
&&&  child 1\cr
&&&\ldots\cr
\ler
&&&  child 1\cr
 PRODUCE-NEXT $\to$ &  1   & $\to$ PRODUCE-NEXT \cr
&&&  child 1\cr
DONE $\from$  &  1   & $\from$ LAST \cr
&&&  transition from child 1{ to child 2}\cr
 PRODUCE-NEXT $\to$ &  2   & $\to$ PRODUCE-FIRST \cr
&&&  child 2\cr
DONE $\from$  &  2   & $\from$ DONE \cr
&&&  child 2\cr
&&&\ldots\cr
\ler 
&&&  child 2\cr
 PRODUCE-NEXT $\to$ &  2   & $\to$ PRODUCE-NEXT \cr
&&&  child 2\cr
LAST $\from$  &  2   & $\from$ LAST \cr
&&&  dormant\cr
}}}
\fi
\caption{The message flow from the viewpoint of a union node. In
    each line, the node receives a message from its parent and sends a
  message to one of its children, or vice verse. The number of the
  involved child is indicated in the second column.}
  \label{fig:message-flow-union}
\end{figure}

\paragraph{Union nodes.}

The message flow of a union node is shown in Figure~\ref
{fig:message-flow-union}, and it is easy to understand.
When receiving a PRODUCE message from its parent, the union node $K$ will
enter exactly one of its two children. Upon returning from a child,
control will pass back to the parent of $K$.
It is obvious the $K$ performs two successive loops over its children.

\begin{figure}
  \centering
\ifjustsyntaxchecking\else
{
  \lineskiplimit=-1,6 \baselineskip
  \baselineskip = 0,55 \baselineskip
  \def\ler{\noalign{\hrule height-2pt}}

  \noindent 
\vbox{  \halign 
  {\strut \tabskip=3pt
  \hfil# & \hfil# \hfil & # \hfil&\qquad # \hfil  \tabskip=0pt plus
  5cm
  \cr
message from/to parent & \textit{ child } &message from/to child&\sta\cr
\noalign{\hrule}
&&&dormant\cr
 PRODUCE-FIRST $\to$ &  1   & $\to$ PRODUCE-FIRST \cr
&&&  working\cr
  &  1   & $\from$ DONE \cr
  \ler
 & 2   & $\to$ PRODUCE-FIRST \cr
&&&  working\cr
DONE $\from$  &  2   & $\from$ DONE \cr
&&&  working\cr
PRODUCE-NEXT $\to$ &  2
\smash{\lower 4mm\rlap{ \ $\left\lgroup\vbox to 16mm{}\right.$}}%
& $\to$ PRODUCE-NEXT \cr
&&&  working\cr
DONE $\from$  &  2   & $\from$ DONE \cr
&&&  working\cr
&&&  \ldots\cr
 PRODUCE-NEXT $\to$ &  2   & $\to$ PRODUCE-NEXT \cr
&&&  working\cr
DONE $\from$  &  2   & $\from$ LAST \cr
&&&  child 2 has finished\cr
 PRODUCE-NEXT $\to$ &  1   & $\to$ PRODUCE-NEXT \cr
&&&  working\cr
  &  1   & $\from$ DONE \cr
  \ler
 & 2   & $\to$ PRODUCE-FIRST \cr
&&&  working\cr
DONE $\from$  &  2   & $\from$ DONE \cr
&&&  working\cr
&&&  \ldots\cr
PRODUCE-NEXT $\to$ &  2
\smash{\raise 3mm\rlap{ \ $\left\lgroup\vbox to 11mm{}\right.$}}%
& $\to$ PRODUCE-NEXT \cr
&&&  working\cr
DONE $\from$  &  2   & $\from$ LAST \cr
&&&  child 2 has finished\cr
 PRODUCE-NEXT $\to$ &  1   & $\to$ PRODUCE-NEXT \cr
&&&working\cr
  &  1   & $\from$ LAST \cr
  \ler
 & 2   & $\to$ PRODUCE-FIRST \cr
&&&  child 1 has finished\cr
DONE $\from$  &  2   & $\from$ DONE \cr
&&&  child 1 has finished\cr
PRODUCE-NEXT $\to$ &
2
\smash{\lower 4,5mm\rlap{ \ $\left\lgroup\vbox to 16mm{}\right.$}}%
& $\to$ PRODUCE-NEXT \cr
&&&  child 1 has finished\cr
DONE $\from$  &  2   & $\from$ DONE \cr
&&&  \ldots\cr
  \ler\cr
 PRODUCE-NEXT $\to$ &  2   & $\to$ PRODUCE-NEXT \cr
&&&  child 1 has finished\cr
LAST $\from$  &  2   & $\from$ LAST \cr
&&&  dormant\cr
}  }}
\fi
\caption{The message flow from the viewpoint of a product node.
  Each inner loop over child~2 is grouped by a bracket. 
  In this example, there are three iterations of the outer loop.
  As in Figure~\ref{fig:message-flow-union}, each line represents one
  operation of the node under consideration, except
  when a  received message from a child results in a message being
  sent to another child: then the operation appears on two consecutive
  lines. The \child\ attribute in the second column identifies also the number of the child
  with whom the message exchange takes place.
}
  \label{fig:message-flow-product}
\end{figure}

\paragraph{Product nodes.}

The message flow of a product node $K$ is shown in Figure~\ref
{fig:message-flow-product}.
The attribute $K.\child$ always stores the number of the child that
was entered from $K$.
The default state is
``working''. If any child has recently sent the LAST message, this is recorded
as the state
``child 1 has finished'' or
``child 2 has finished''.
One can see that
$K$ implements a nested loop.
%

When receiving a PRODUCE message from its parent, $K$ 
will
 enter the second child or both children
 before passing control back to the parent.
 The first child will only be visited on the first activation
from the parent
with the message PRODUCE-FIRST,  or after the inner loop (of the second
child) has been
exhausted on the previous visit, which is indicated by
the state
``child 2 has finished''.
After the visiting the first child, the loop over the second child
will be initialized
with a PRODUCE-FIRST
 message.

The analysis of the algorithm is a straightforward modification of 
the analysis in Section~\ref{listing}.
Recall that we defined a
{\cwstree} as a subtree $E$ of the expression DAG that contains
both children of every product node
in~$E$
and exactly one child of every union node
in~$E$.
A \emph{partial \wstree} is defined similarly,
except that
 a product node may also have
just one child in~$E$.

\begin{proposition}
  If a node $K$ receives a request from a parent,
  Algorithm ENUM3 will visit the nodes of
  partial \wstree\ with root $K$ before replying to the parent.
 \qed
\end{proposition}

The set of visited nodes is actually the same as those nodes that are visited
by a proper visit in Algorithm ENUM1.

A partial \wstree\ can easily be extended into a (complete) \wstree.
Therefore, by Lemma~\ref{structured-trees}(3), a partial
\wstree\ whose root is associated to the vertex set~$A$ contains
$O(|A|)$ nodes in total.
We conclude:

\begin{theorem}
    Algorithm ENUM3
  enumerates
   the \mds s of a
   tree with linear delay, after linear setup time.
   After the last solution, the algorithm terminates in constant time.
\end{theorem}


\vfill\eject
\ifjustsyntaxchecking\else
\ifARXIV
\section{Overview of Notations}
\else
\section{Overview of Notations (NOT FOR INCLUSION IN THE JOURNAL
  PUBLICATION)}
\fi

\begin{itemize}[noitemsep]
\item 
$T$ = a tree $T=(V,E)$
\item 
a graph $G=(V,E)$
\item $n=|V|$ = number of vertices
\item 
$D\subseteq V$  a dominating set
\item $A\subseteq V$ a subtree 
\item \textbf{G}ood. $\ne$ graph $G$
\item \textbf{S}elf 
\item \textbf{L}acking
\item \textbf{d}ominated
\item \textbf{p}rivate
\item \textbf{f}ree
\item  subtrees $A_1$, $A_2$, $B$ combined into a tree $C$
\item vector $v=(G,S,L,d,p,f)$
\item $\bar M(G,S,L,d,p,f)=G+S+d+p$ = \#MDS
\item with root $r$, and $s$
\item special vertices $a$ and $b$ in the star of snowflakes 
\item general vertices $a$ and $b$ 
\item total number $M(T)$ 
\item $k$ = number of snowflakes
\item  $\mathrm{RT}_{13k+1}$ record trees
\item $M_n$ = max \# MDS
\item $\mathcal{V}_n
$ = set of 6-vectors for trees of size $n$
\item $v_i$ = individual 6-vectors, vertices of $P$
\item $v_0 = (0,1,0,0,0,1)$, starting vector
\item $\preceq,\succeq$ majorization
\item $v \star v'$ for individual vectors, $w, w'$
\item $V \circ V'$ for sets of vectors
\item $P,Q$ sets of vectors, $P$ ``polytope'', $Q$ discrete set
\item $\lambda,\lambda^*$ = growth rate
\item $\mu_i,\nu_j$ coefficients for convex combination
\item $\hull( P)$ majorized hull
\item $\#\hull(P)$ number of its generating vertices = nonmajorized
  vertices (used only once)
\item $\hullp (P)$ majorized convex hull
\item $\#\hullp(P)$ number of its extreme vertices number (used only once)
\item $\mathbf{X}=\mathbf{X}(T)$
  Expression Dag 
\item $K,K',K_2,K_2$ nodes in the expression DAG, also in the context of the
  program,
as a record or \emph{object}
\item $R(K),R(K_1)\subseteq 2^V$ = the node subsets \emph{represented}
  by $K$
\item $k$ iterations in a generator loop
\item $C_1,C_2$ number of solutions represented by child 1/2
\item $t_1,t_2,t,t'$ average time for enumeration
\item $k=a \log_2 n$ number of stars in the chain of star clusters example
\item $E$ subgraph of visited nodes, \wstree
\item $p$ number of visited product nodes
\end{itemize}

\fi
\fi

\end{document}